\newif\ifOLDimc
\newif\ifimc
\newif\ifietf
\newif\ifconext

\imctrue
\newcommand{\PaperTitle}{Rooting Out Incorrect RIPE Atlas Probe Geolocations}

\ifOLDimc
\documentclass[10pt,sigconf,letterpaper,nonacm]{acmart}
\fi
\ifimc
\documentclass[10pt,sigconf,letterpaper,nonacm]{acmart}
\fi
\ifconext
\documentclass[acmsmall,anonymous,review]{acmart}
\fi
\usepackage{colortbl, xcolor}
\usepackage{booktabs, multirow, xcolor, caption}
\usepackage{graphicx}
\usepackage[labelformat=simple]{subcaption}
\usepackage{xspace}
\usepackage{supertabular}
\usepackage{url}
\usepackage{hyperref}
\usepackage{float}
\usepackage{pdflscape}
\newcommand{\todo}[1]{}
\renewcommand{\todo}[1]{{\color{blue} TODO:  {#1}}}

\newcommand{\liz}[1]{}
\renewcommand{\liz}[1]{{\color{orange} LIZ:  {#1}}}

\definecolor{blush}{rgb}{0.87, 0.36, 0.51}
\newcommand{\changed}[1]{}
\renewcommand{\changed}[1]{{\color{blush}{#1}}}

\newcommand{\ark}{Ark\xspace}

\newcommand{\presentday}{April 2026\xspace}
\newcommand{\totalallyears}{731\xspace}
\newcommand{\numviolatingtoday}{155\xspace}

\newcommand{\atlas}{RIPE Atlas\xspace}

\newcommand{\sol}{speed of Internet\xspace}

\usepackage[tableposition=below]{caption}

\definecolor{chicagomaroon}{rgb}{0.5, 0.0, 0.0}
\iftrue
\newcommand{\katherine}[1]{\textcolor{chicagomaroon}{\emph{#1}}}
\newcommand{\sumanth}[1]{\textcolor{violet}{\emph{#1 --SR}}}
\newcommand{\alisha}[1]{\textcolor{green}{\emph{#1 -- AU}}}
\newcommand{\ben}[1]{\textcolor{purple}{\emph{#1 --BD}}}
\else
\newcommand{\katherine}[1]{}
\newcommand{\ben}[1]{}
\newcommand{\sumanth}[1]{}
\newcommand{\alisha}[1]{}
\newcommand{\liz}[1]{}
\fi
\begin{document}

\title{\PaperTitle}

%\author{Paper \#203, 6 pages body, 11 pages total}
\author{%
\small
Katherine Izhikevich\textsuperscript{1},
Ben Du\textsuperscript{2},
Manda Tran\textsuperscript{2},
Sumanth Rao\textsuperscript{1},
Alisha Ukani\textsuperscript{1},
Ray Bellis\textsuperscript{3},
Liz Izhikevich\textsuperscript{2} \\
\textsuperscript{1}UC San Diego 
\textsuperscript{2}UC Los Angeles
\textsuperscript{3}Internet Systems Consortium
}
% \author{Katherine Izhikevich}
% \affiliation{%
%   \institution{UC San Diego}
%   \country{}
% }
% \email{katherine@ucsd.edu}

% \author{Ben Du}
% \affiliation{%
%   \institution{UC Los Angeles}
%   \country{}
% }
% \email{c4du@ucla.edu}

% \author{Manda Tran}
% \affiliation{%
%   \institution{UC Los Angeles}
%   \country{}
% }
% \email{mandat@ucla.edu}

% \author{Sumanth Rao}
% \affiliation{%
%   \institution{UC San Diego}
%   \country{}
% }
% \email{svrao@ucsd.edu}

% \author{Alisha Ukani}
% \affiliation{%
%   \institution{UC San Diego}
%   \country{}
% }
% \email{aukani@ucsd.edu}

% \author{Ray Bellis}
% \affiliation{%
%   \institution{Internet Systems Consortium}
%   \country{}
% }
% \email{ray@isc.org}

% \author{Liz Izhikevich}
% \affiliation{%
%   \institution{UC Los Angeles}
%   \country{}
% }
% \email{lizhikev@ucla.edu}

\ifconext
    %\input{gameplan}
    %\newpage
    \begin{abstract}
Geolocation plays a critical role in understanding the Internet. 
In this work, we analyze and fix operator-misreported geolocation.
Using DNS root servers to detect speed of Internet violations, we conservatively infer that at least 2\% of vantage points in the largest community-vantage point collection, RIPE Atlas, were not located at their operator-reported geolocation between January 2019 and \presentday. 
%Furthermore, the problem is escalating: within the past seven years, the number of probes likely reporting the wrong location has increased over five-fold. 
To increase the accuracy of future studies that use operator-reported geolocation, we open source our simple methodology, implement a reporting campaign of violating probes within RIPE, work with operators to fix the misreported geolocations, and release a continually updated dataset of RIPE vantage points that likely misreport geolocation.
\end{abstract}
    \maketitle
    \vspace{-3mm}

\section{Introduction}

IP address geolocation plays a central role in networking and security: 
understanding the Internet's topology to improve last mile latency~\cite{lastmilelatency, lastmilecongestion}, filtering traffic to prevent cyber attacks~\cite{wan20origins}, and identifying malicious activity~\cite{izhikevich2024using,dnsintercept}, all rely on geolocation. 
To geolocate an IP address, methodologies often rely upon a global, human-reported set of vantage points (i.e., probes) to measure from and verify accuracy against. 
For example, geolocation based on latency measurements often require a network of probes, each with operator-attested geolocation, to be located near every target to accurately estimate every target's geolocation~\cite{hu2012towards,gueye2004constraint,katz2006towards,wong2007octant,calder2013mapping,fanou2018exploring}.

%Thus, such a set of vantage points need be geographically distributed, relying on operators to attest to the location of each probe.
% Yet, despite geolocation's importance for nearly two decades, we still have no method to geolocate IPs with 100\% accuracy~\cite{darwich2023replication}.

%latency is often distorted due to a diversity of routing paths, network types (e.g., optical fiber, copper wire, satellite, mobile), and congestion. 
% Geolocating IPs is non-trivial, in part because it requires a global set of vantage points.
% \changed{For example, geolocation methods that rely on accurate latency measurements require that a vantage point exist nearby every target~\cite{hu2012towards,gueye2004constraint,katz2006towards,wong2007octant}.}
% Verifying the global accuracy of a geolocation methodology requires that a large-enough set of vantage points with already-known geolocations exists to compare against.
%Verifying geolocation's global accuracy requires a large-enough set of distributed IPs with already-known geolocations to compare against (i.e.,  a ground truth dataset). 

While necessary, relying on human operators to report accurate probe geolocation can introduce human error.
For example, if an operator moves their probe, but fails to update its location, geolocation methodologies relying on the outdated information risk inaccurate conclusions.
To account for incorrect geolocation, a few geolocation studies take an additional precaution to filter out probes reporting unlikely locations~\cite{darwich2023replication, gharaibeh2017look}.
% ,fanou2018exploring,calder2013mapping}.
%Since filtering erronous vantage points is not the broader intention of geolocation studies,
%existing filtering methodologies often require high bandwidth~\cite{darwich2023replication} and risk filtering accurately geolocated probes~\cite{gharaibeh2017look,darwich2023replication}.
However, where misreported probes occur, the magnitude of their error, and whether they are corrected remains poorly understood.
% Furthermore, and most critically, prior work stops short of addressing the problem at its source.
As operator-reported geolocations continue to serve as a foundation for many studies~\cite{bajpai2015lessons,holterbach2015quantifying,iodice2019periodic,fanou2015diversity,corneo2021surrounded,gigis2017characterizing,sukhov2014evaluating,davisson2021reassessing,candela2019using}, it is critical that we understand the successes and pitfalls of our community-contributed datasets and work towards an accurate ground truth.
\looseness=-1

In this work, we provide the most in-depth analysis to date of operator misreported geolocation, including where violations occur, what causes them, and how operators correct inaccuracies after we report their probes.
First, we apply a simple methodology that uses speed-of-Internet calculations between probes and DNS root servers to conservatively test whether operator-reported probe locations are physically possible.
%First, we apply a simple methodology that conservatively infers the set of probes that do not respond from their operator-reported geolocation.
%The methodology (1) uses speed of Internet calculations to derive a minimum bound of expected latency between community-provided vantage points and DNS root servers and 
%(2) applies the minimum bound to infer whether the community-reported geolocations are physically possible.
We apply this methodology to RIPE Atlas~\cite{ripeatlas}, the largest and most geographically distributed community vantage-point collection, using 2.1K~anycasted DNS root servers between January~2019 and \presentday.
Our methodology infers a lower bound of \totalallyears probes (2.4\%) are likely not responding from their reported location, as doing so would exceed the physical limits of the speed of light in Internet fiber.
We show that this seemingly small error rate has cascaded into hundreds of incorrect city-level geolocations in a prior study and thousands in scheduled \atlas measurements.
\looseness=-1

Errors in misreported geolocation are substantial; roughly 50\% of violating probes likely responded at least 1,000~kilometers away from their reported location. 
We consider how inaccuracies prevail over time, finding that operators are slow to update geolocation and that the number of violating probes is increasing. The majority (52\%) of geolocations are updated after the probe violates the speed of Internet for at least 13~weeks. 
Furthermore, the proportion of violating probes increases over time, highlighting that this problem is not diminishing.

We correct misreported geolocation at the source by working with RIPE Atlas to report physically impossible 
geolocations, prompting over 90\% of responding operators to update their probe locations.
To support this effort, we build a reporting pipeline that continually identifies probes that violate \sol and use operator responses to validate our methodology.
We open-source our methodology so researchers can identify physically impossible operator-reported geolocations in other settings.
Many studies rely on operator-reported geolocation~\cite{surr_by_clouds,NREN_africa,fanou2017four,darwich2023replication,dnsintercept,lastmilecongestion,lastmilelatency}, and our approach provides a lightweight method to root out physically impossible geolocations to more accurately measure the Internet.
\enlargethispage{2\baselineskip}
    \section{Background and Related Work}\label{sec:background}
Many studies rely on RIPE Atlas to develop geolocation methodologies and understand the Internet, making them dependent on the accuracy of operator-reported locations.
%This motivates us to find vantage points with misreported geolocations, more deeply understand the inaccuracies, and lower the barrier for researchers to validate probes.

%\todo{liz's proposal: accurate geolocation is important. see ripe and all the things that geolocation is important for that. however, prior studies have hinted that operator reported geolocation isn't always accurate. so to validate, what prior work has done is used a set of vantage points and speed of internet tests to see if reported locations are physically possible. However, they often use the set of untrustworthy set of vantage points themselves (ripe nodes themselves), which leads to issues. we take a different approach, we identify a publicaly available set of distributed nodes whose geolocation we also trust: DNS root servers. now background on root serveres...and why  we trust them (but also do verify their location later too, just in case). Now, root servers do present a problem: they are anycasted, so you dont actulaly know who is responding.  but we find a fix for that, BIND queries.   }

\vspace{3pt}
\noindent
\textbf{RIPE Atlas.}\quad
RIPE Atlas is a global measurement platform with 14,405 active vantage points across 183 countries and 4,475 ASes as of \presentday~\cite{ripeatlas}.
It includes volunteer-hosted probes and organization-hosted anchors; anchors are considered more ``accurate''~\cite{davisson2021reassessing}  because they require RIPE-approved hardware. 
Unless stated otherwise, we refer to both as RIPE Atlas probes.
%RIPE Atlas vantage points include two types: anchors (hosted by organizations) and probes (hosted by volunteers).
%Anchors are considered more ``accurate''~\cite{davisson2021reassessing} because hosts must purchase RIPE-approved hardware and either build the anchor themselves or install it on a high-performance virtual machine.
%We refer to both anchors and probes as \atlas probes unless stated otherwise.
%The RIPE NCC provides a convenient interface for participants to manage their probes: after the probe is connected, the probe host can see their probe metadata (e.g. prefix and AS number) on their RIPE Atlas dashboard and report the physical location of their probe by placing a pin on a map. 
Operators report their probe’s physical location by placing a pin on a map, and must manually update this location if the probe moves. 
The reported probe location will be skewed at most by 1 kilometer to protect the probe host's privacy.~\footnote{We observed the same anonymization for the 6 \atlas probes we control.} When a probe becomes disconnected, the RIPE Atlas platform will regularly send the host email notifications to remind them of the disconnection. 
Critically, if the probe is physically moved, the operator must manually update their reported location. 
The RIPE platform allows other users to dispute a given probe's geolocation, although prior to our notification campaign, only 32~probe locations were disputed.

\vspace{3pt}
\noindent
\textbf{Research Using RIPE Atlas.}\quad
\label{sub:sec:research_related}
The distributed architecture of the \atlas platform has proven useful in many research areas, especially in Internet infrastructure geolocation~\cite{dnsintercept, lastmilecongestion, lastmilelatency,surr_by_clouds,NREN_africa,fanou2017four}. 
Scheitle et al.~\cite{hloc} used \atlas probes to verify geolocation inferred from location hints embedded in reverse DNS records. 
Weinberg et al.~\cite{proxylie} used \atlas probes to measure the latency to 2K proxy and VPN providers, and found at least 1/3 of VPNs falsified their advertised geolocation.
Fanou et al.~\cite{fanou2018exploring} use speed-of-light checks to identify that 20\% of probed Google cache and DNS resolver IPs are wrongly geolocated.
Darwich et al.~\cite{darwich2023replication} used RIPE Atlas to replicate the results of previous geolocation works. 
Gharaibeh et al.~\cite{gharaibeh2017look} used \atlas probes to validate four geolocation databases. 
%probes and anchors as vantage points and targets 
RIPE themselves use \atlas probes to power a public IP geolocation tool, IPmap~\cite{ripeipmap,ipmapeval}.

% Given their known locations, RIPE Atlas probes can serve as vantage points capable of sending traffic to target IP addresses. The RIPE NCC has created IPmap, a public IP geolocation tool \cite{ripeipmap}. 
% One of its components, the IPmap single-radius engine, uses RIPE Atlas probes around the world to ping the target IP address and estimates the geolocation of the target based on the lowest obtained latency \cite{ipmapeval}. 

%In addition, RIPE Atlas can also be used as measurement targets to evaluate the accuracy of geolocation techniques.  %but were not able to achieve the same level of accuracy or coverage. 

%NOT ALL METHODS USE LATENCY-BASED METRICS!! please keep "often" or "most" do not change!

Many geolocation studies using RIPE Atlas probes are sensitive to probe-location accuracy because latency-based methods often assume that nearby probes provide evidence for nearby targets.
For instance, if RIPE IPmap sees a 0.2~ms RTT to a mislocated probe, it may wrongly assign the target IP to that incorrect location.
Probes with incorrect reported locations have a larger impact on results where RIPE Atlas probes are sparsely deployed.
% \todo{Candela et al. \cite{atlasforgeo} studied the maximum theoretically achievable geolocation accuracy for RIPE Atlas-based techniques, and found that regions with limited probes, such as Africa and Asia, have lower maximum accuracy.}
Trammell and Kühlewind~\cite{luckgeo} also found that accuracy depends upon selecting a vantage point close to the target. 
%Such luck would be defeated in a situation where a probe reporting an incorrect location is selected.

Given the importance of accurate geolocation, prior work filters unlikely RIPE Atlas probe locations, but stops short of understanding or correcting misreported geolocation at the source. 
Gharaibeh et al.~\cite{gharaibeh2017look} identify 24 questionable probes using basic sanity checks, while Darwich et al.~\cite{darwich2023replication} use speed-of-Internet violations among RIPE Atlas anchors and probes themselves to filter 9 of 723 anchors and 96 of 10K probes; we further describe their methodologies in Appendix~\ref{app:prior-work}. 
We introduce a methodology that uses DNS root servers as the reference points for detecting physically impossible RIPE Atlas probe locations.

%Their reliably maintained locations provide a stronger ground truth for validating RIPE Atlas probe geolocation.}

%original longer version
% To the best of our knowledge, few works validate the correctness of RIPE Atlas probe locations. 
% Gharaibeh et al.~\cite{gharaibeh2017look} in 2017 were the first to include a preliminary ``sanity check'' of RIPE Atlas probe locations. 
% They considered the reported locations untrustworthy if they had default country coordinates (typically geographic centers) or if two probes were connected to the same router, but reported locations more than 100km apart. 
% They identified 24 probes with potentially incorrect geolocation. 
% In 2020, Darwich et al.~\cite{darwich2023replication} introduced a recursive-like methodology, described in Section~\ref{sub:sub:sec:comp_prior}, to identify probes that violated the speed-of-Internet constraint, filtering 9/723~anchors and 96/10K probes.
% Notably, both methodologies identify groups of probes that contain at least one geolocation violation, but they do not pinpoint which specific probe within the group is responsible.
% Thus, both methodologies risk filtering probes that accurately report their geolocation. 

%keep the darwich and gharaibeh sentences in the same paragraph, they have a topic sentence tying them together
%keep how our work is different a sep. paragraph

\vspace{3pt}
\noindent
\textbf{DNS Root Servers.}\quad
%The Domain Name System (DNS) is a hierarchical system whose core function is to map hostnames to their corresponding IP addresses. At the top level of the hierarchy. 
Domain Name System (DNS) root servers provide a globally distributed, operationally managed reference infrastructure for our measurements.
The DNS root zone is served by 13 root servers, A through M~\cite{ianarootservers},
whose operators deploy thousands of anycasted instances worldwide~\cite{rootserversorg}.
%As a fundamental Internet infrastructure, root servers use anycast to improve resilience against outages and other service disruptions~\cite{ditlanycast}, and as of 2025, there are thousands of DNS root server instances hosted by 12 root server operators (RSOs) in the world~\cite{rootserversorg}. 
%Some root  deploy all server instances themselves, while others partner with network operators to host instances in underserved regions. Regardless of the deployment model, 
Operating a root server instance requires meeting strict technical and administrative criteria that ensure the root server operators retains reliable control, consistent performance, and secure management~\cite{hostfroot,hostiroot,hostkroot,hostlroot}. These requirements are designed to uphold the service expectations defined by the Root Server System Advisory Committee (RSSAC)~\cite{rssac001}. 

We use DNS root-server instances to validate the reported geolocations of RIPE Atlas probes.
Because root servers are anycasted, multiple instances can respond from the same IP address; we therefore use standardized \texttt{hostname.bind} responses to identify which instance answered each query~\cite{RFC4892}.
Critically, RIPE Atlas already collects measurements between probes and DNS root servers, allowing us to validate reported probe locations using existing historical data rather than issuing large volumes of new measurements~\cite{darwich2023replication}.
%However, because root servers are anycasted, different root server instances respond from the same IP address, making it difficult to determine which specific instance answered a query. 
%To disambiguate which instance responded to the query and more broadly support operational diagnostics, DNS operators standardized the use of \texttt{hostname.bind} queries, in which a responding DNS server returns an identifying string for the instance handling the request~\cite{RFC4892}.

% the location of the responding instance is not immediately apparent.
% We address this challenge in \S\ref{sec:meth} by issuing \texttt{hostname.bind} queries, which reveal the geolocation of the specific server that responds.

\vspace{3pt}
\noindent
\textbf{Precautions When Using DNS Root Servers.}\quad
%We introduce a few studies that are relevant to the context of this paper. 
Prior work on DNS root servers motivates several precautions in our methodology.
%Prior work uses \texttt{hostname.bind} queries to analyze the inefficiencies of anycast routing toward DNS root servers. 
Li et al.~\cite{anycastperformance} measured the RTTs from RIPE Atlas probes to C, D, and K roots and found that more than 2/3 queries were directed to a root instance that is not the closest to the Atlas probe. 
Similarly, Koch et al.~\cite{anycastincontext} found that root servers with larger anycast deployments are more likely to experience inflated latencies. 
% lthough the latencies from Atlas probes to root servers may not be minimal, such inflation does not introduce false positives into our inference of Atlas probes with incorrectly reported geolocation.
Other studies have analyzed the manipulation of DNS root server responses. In 2016, Jones et al.~\cite{rootmanipulate2016} compared RTTs from RIPE Atlas probes obtained by pings and \texttt{hostname.bind} and found only 11 instances of B root server response manipulations and none for L root. They also anecdotally identified one RIPE Atlas probe in New York that was mis-reported to be in Switzerland. In 2020, Wei and Heidemann~\cite{wei2020whacamoleyearsdnsspoofing} identified two types of DNS response manipulators: 1) \textit{Overt spoofers} who responds to \texttt{hostname.bind} queries with their own identifiers;
% and almost responds faster than actual root server instances. 
% Queries responses from this type of spoofers have lower RTTs and may cause false-positives in our methodology, and thus we filter out these queries in \S\ref{sec:meth}.
2) \textit{Covert delayers} who respond with identifiers that are the same as the root server instances. 
% These delayers cannot reduce the response latency and therefore does not cause false-positives in our inference, and thus we do not filter out these responses.
We minimize the risk of inflated latencies and these manipulators in Section~\ref{sec:meth}.

%Our approach identifies 4~times more violating probes while using 9~times less bandwidth (see Section~\ref{sub:sec:eval}) than prior work. 

\looseness=-1
%liz: included in into
%\todo{1. SHould I include research using Atlas that are not sensitive to location correctness, like topology studies -- which ISPs have the most last mile latency\cite{lastmilelatency, lastmilecongestion}, and which ISPs do DNS interception \cite{dnsintercept}}

%liz: related reads well! motivation is good, we will keep related work as second section
%\todo{2. This also kinda feels like a intro/motivation...}
    \section{Methodology}
% Studies can use probes in \atlas as measurement sources and targets.
% \todo{is it still two methodologies? or now its one?}
In this section we outline our methodology to measure whether probes' self-reported locations are physically impossible.
We measure the latency between RIPE Atlas probes and DNS root servers using \texttt{hostname.bind} queries.
We rely on simple physics to infer whether the measured latency between our measurement source and the destination's reported location exceeds the speed at which light travels in a fiber optic cable.
\looseness=-1
% We measure latencies from RIPE Atlas probes to DNS Root server instances using \texttt{hostname.bind} queries that were originally designed to identify the responding DNS server for debugging and operational diagnostics~\cite{dnsintercept}.
\label{sec:meth}
%\subsection{Outbound DNS Queries}

\vspace{3pt}
\noindent
\textbf{Obtaining \texttt{hostname.bind} Responses.}\quad
RIPE Atlas operates a suite of built-in measurements that each probe conducts automatically.
As part of these measurements, every connected probe issues a \texttt{hostname.bind} query to each of the 13 DNS root server identities every 240 seconds~\cite{atlasbuiltin}.
We obtain historical \texttt{hostname.bind} responses collected by all connected probes between January 2019 and \presentday from RIPE Atlas' historical measurements database~\cite{ripe-atlas-api}.

\vspace{3pt}
\noindent
\textbf{Sanitizing \texttt{hostname.bind} Responses.}\quad
Wei and Heidemann~\cite{wei2020whacamoleyearsdnsspoofing} identified several types of DNS manipulators that may respond to \texttt{hostname.bind} queries in place of the actual root server instances (\S\ref{sec:background}). 
Critically, they explained intercepting ISPs respond to \texttt{hostname.bind} queries with naming conventions that do not follow the naming scheme of A--M roots.
%They further explained that the \textit{overt spoofers} are mostly ISPs who intercept DNS traffic for purposes such as content moderation and censorship, and most responds faster than actual root server instances. Queries responses from \textit{overt spoofers} may have lower RTTs and thus cause false-positives in our methodology by incorrectly considering it as being lower than our calculated \textit{theoretical absolute minimum RTT}.
To account for this phenomenon, we remove any \texttt{hostname.bind} responses that do not match the naming conventions of A -- M roots. 
Covert delayers, who intercept the query and therefore cause inflated latencies, do not affect our results, as we explain in the following section.
% Wei and Heidemann~\cite{wei2020whacamoleyearsdnsspoofing} also identified few \textit{covert delayers} who intercept the query and eventually forward the responses. These actors may cause inflated latencies in the responses but will not cause false positives in our methodology and thus we process these responses as usual.
% We remove those spoofed responses using Wei and Heidemann's~\cite{wei2020whacamoleyearsdnsspoofing} findings: they found that the intercepting ISPs respond to \texttt{hostname.bind} queries with unique identifiers of their own recursive resolvers which follow their own naming conventions different from that of the root server instances. We create regular expressions that follows the naming conventions of A -- M roots and remove any \texttt{hostname.bind} responses that do not match those regular expressions. Wei and Heidemann~\cite{wei2020whacamoleyearsdnsspoofing} also identified few \textit{covert delayers} who intercept the query and eventually forward the responses. These actors may cause inflated latencies in the responses but will not cause false positives in our methodology and thus we process these responses as usual.

\vspace{3pt}
\noindent
\textbf{Extracting Latency Values.}\quad
Although \texttt{hostname.bind} was originally intended for operational diagnostics, we repurpose these measurements to infer latency between RIPE Atlas probes and individual DNS root server instances.
When a RIPE Atlas probe sends a \texttt{hostname.bind} query, the responding DNS root server instance returns a hostname identifying itself, and the probe simultaneously measures the round-trip time (RTT) between transmitting the query and receiving the response.
% The returned hostname encodes the location of the responding root server instance either by (1) matching an instance listed in \texttt{root-servers.org}, which provides the expected latitude and longitude of that instance as reported by the Root Server Operator (RSO), or (2) embedding an IATA airport code within the hostname.
% Combining the \texttt{root-servers.org} instance registry with IATA code extraction allows us to geolocate 98\% of observed responding hostnames in \presentday.
After parsing the \texttt{hostname.bind} responses, the resulting dataset consists of the initiating probe ID, DNS root server identity (A–M), the hostname of the responding anycast instance, and the RTT measured by the probe.

We sample measurements three times per day (06:00, 12:00, and 18:00 UTC) on the first day of each month, beginning on January 1, 2019.
Then, for each \texttt{<probe,instance>} pair, we retain the minimum observed RTT to each root letter per day as the representative latency.
We use minimum RTTs as the representative RTT because the speed of light in optical fiber or air imposes a hard lower bound.
Unlike averaged RTTs, which can be inflated by congestion or queuing effects, the minimum RTT reflects the fastest observed path and cannot be artificially reduced.

\looseness=-1

\vspace{3pt}
\noindent
\textbf{Estimating Distance Between Source and Destination.}\quad
The \texttt{hostname.bind} responses encode the location of the responding root server instance by matching an instance listed in \texttt{root-servers.org}, which provides the expected latitude and longitude of that instance as reported by the Root Server Operator (RSO).
For hostnames not explicitly defined in the \texttt{root-servers.org} dataset (7\%), we infer their location by matching the site prefix from the dataset. 
If all listed instances with a matching prefix (e.g., \texttt{nnn1-ams[4-6]}) share the same coordinates, we assume an unlisted instance matching the same format (e.g., \texttt{nnn1-ams1}) is co-located and assign it those coordinates accordingly.
Combining these techniques we are able to geolocate the observed responding hostnames for 94\% of our measurements between January 2019 and \presentday, removing the remaining 6\% from our study.
\looseness=-1

To help derive the theoretical minimum latency between a vantage point and probe, we estimate the minimum Haversine distance (``as-the-crow-flies''), hereafter referred to as $H_d$, between the reported coordinates of every \atlas probe and every DNS root server instance. 
By leveraging \atlas provided geolocation snapshots, we ensure we are using the reported location of the probe from the date of each measurement.
Recognizing that root server coordinates are often approximated (e.g., city-center or airport locations), we incorporate a substantial 100~km error radius,\footnote{The vast majority of cities in the world are significantly less than 100~km in radius.} so our analysis identifies only significant geolocation discrepancies.

\vspace{3pt}
\noindent
\textbf{Deriving Speed of Internet Theoretical RTTs.}\quad
After estimating the minimum distance between each probe and server ($H_d$), we convert it to a minimum theoretical RTT.
We calculate the theoretical \textit{absolute minimum} RTT if the pings were sent at the speed of light in optical fiber, $H_d/\frac{2}{3}c$~\cite{chaudhry2022optical}. 
We refer to the speed of Internet as SOI.
% For Starlink-hosted probes, we conservatively estimate the theoretical \textit{absolute minimum} RTT as $Haversine\_distance/c$, where $c$ is the speed of light in a vacuum, as an extra precaution for being conservative and estimating the lower bound.
%\footnote{Note that if a Starlink probe is mobile (e.g., on a ship~\cite{hitchhiking}) and thus has no singular location, such a probe is not fit for a platform that assumes stationary locations. At any rate, if the probe violates the speed of Internet, we flag the probe as a violation.}
% Otherwise,

\vspace{3pt}
\noindent
\textbf{Identifying Violating Probes.}\quad
We identify a \atlas probe as ``violating'' (i.e., we believe the probe is not in its operator-reported geolocation) if any \atlas probe measured latency is lower than the theoretical minimum latency from the self-reported geolocation to at least two DNS root servers operated by distinct root-server operators (a threshold selected to mitigate false positives, as discussed below).
For DNS root server historical data, we take note of when within the past 5~years the RTT violation happened, if not presently occurring. 
% Furthermore, since RIPE halted their built-in measurements to their central servers, we continuously check whether the last set of violating probes from September 2024 have updated their location or disconnected from the platform. If not, we label them as still violating.
% In Section~\ref{sec:disputes}, we compare our predictions with operator-verified updates =

\vspace{3pt}
\noindent
\textbf{Precautions to Minimize False Positives.}\quad \label{sec:min_fp}
Minimizing false positives requires that DNS root servers report accurate locations and that probes can freely communicate with them.  
We consulted a root-server operator who noted that RSOs are generally highly reliable in tracking where their nodes are installed.  
Nevertheless, we take several precautions to minimize false positives.

First, we verify that the DNS root servers we rely on do not themselves violate SOI.
Specifically, we use CAIDA's \ark{}~\cite{caidark}, a set of 294 globally distributed vantage points whose exact latitude and longitude we confirmed directly with CAIDA operators who themselves deploy the servers.   
These nodes span 194 autonomous systems, 218 cities, 68 countries, and 6 continents; a map of Ark nodes is available in~\cite{caidark}.  
Using this independent vantage set, we issue \texttt{hostname.bind} queries and measure RTTs to DNS root servers, finding only one I-root instance appears to violate SOI.
We flag it for further analysis, finding additional evidence in Section~\ref{sec:ablation} that it is likely not in its reported place.
For all other root instances, using CAIDA's \ark{} provides an external sanity check that root-server locations and paths are consistent with physical latency constraints.

Second, we require that a probe violate SOI constraints with at least two DNS root servers operated by distinct root-servers (\texttt{a}--\texttt{m}).  
This mitigates the risk that an error or misconfiguration by a single operator could incorrectly trigger a violation.
Finally, we confirm that none of the violating probes originate from ASes known to perform DNS filtering, following Wei and Heidemann~\cite{wei2020whacamoleyearsdnsspoofing}.  
These precautions collectively reduce false positives, albeit at the cost of increasing false negatives.  

Third, we remove Starlink probes from analysis.
Satellite links can achieve lower latency than terrestrial fiber (e.g., via inter-satellite lasers)~\cite{chaudhry2022optical}.
As a result, Starlink probes may legitimately exceed the terrestrially-defined SOI constraint, so we do not classify them as violators.
We identify these probes via ASN 14593 and exclude them.\footnote{Less than 1\% of probes in April 2026 are announced by the Starlink ASN.}

% Third, we remove Starlink probes from analysis.
% Starlink satellites can achieve lower latency than terrestrial networks because packets traverse links with higher signal propagation speeds: inter-satellite laser links operate near the speed of light in vacuum, and wireless links through air are faster than optical signals in fiber, where propagation is slowed by the refractive index of glass~\cite{chaudhry2022optical}. As a result, Starlink probes may legitimately exceed the SOI constraint, so we do not classify them as violators.\footnote{Less than 1\% of probes in April 2026 are announced by the Starlink ASN.}
% To find probes hosted over Starlink, we look for probes whose IPv4 or IPv6 addresses are announced by Starlink's autonomous system (ASN\,14593)~\cite{hitchhiking}.

We discuss the implications and limitations of this methodology in Section~\ref{sec:limitations}.

    \section{Results}
\label{res}

Between January~2019 and \presentday, \totalallyears~unique probes (2.4\%) violate \sol (SOI).
As of \presentday, \numviolatingtoday~probes are in violation of \sol.
Violating probes are geographically widespread, often responding from thousands of kilometers away from their operator-reported geolocation.
Slow operator response times exacerbates the issue: 50\% of probes violate for over 13~weeks until operators update. We work with \atlas to report inaccurate geolocation and find that operators do update their probe locations, in some cases even moving them across continents.
\begin{figure}[H]
    \centering
    \includegraphics[width=\linewidth]{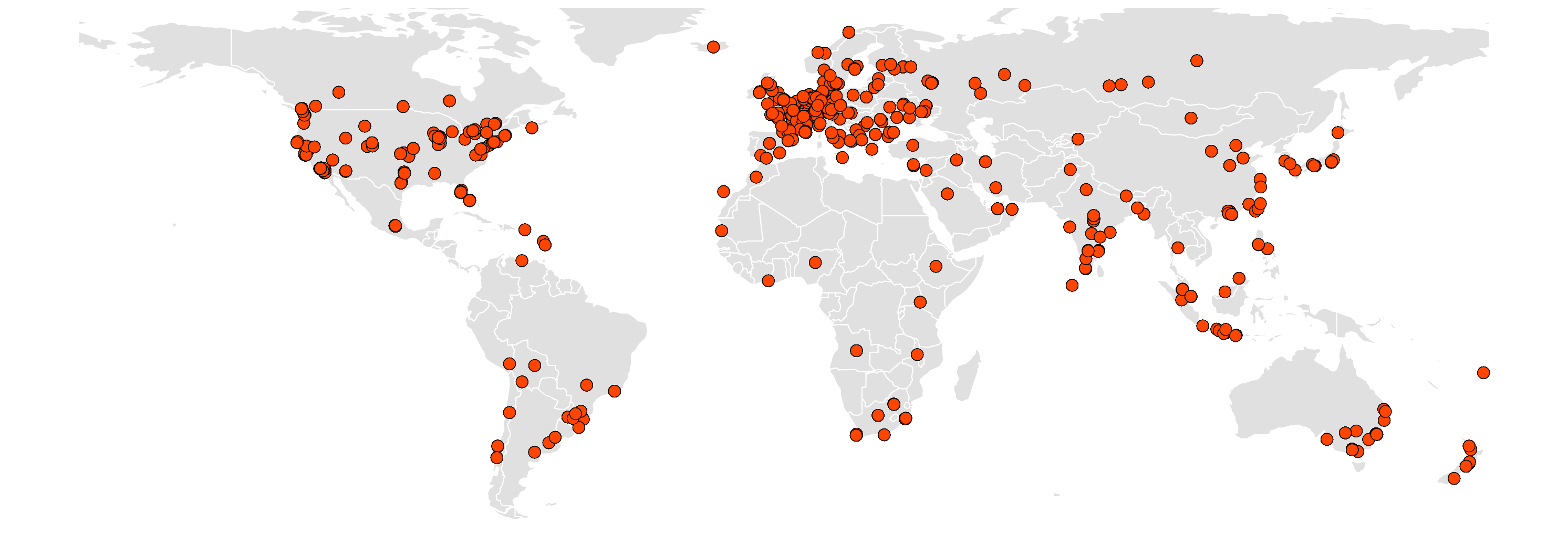}
    \caption{\textbf{Reported geolocations of probes violating SOI between January 2019 and \presentday}---\textnormal{Violating probes are geographically distributed.}}
    \label{fig:violating_map}
\end{figure}
%\vskip 0.1cm
% \vspace{-5mm}
\subsection{Where and When Probes Violate}
Violating probes exist all over the world. 
In Figure~\ref{fig:violating_map}, we plot the reported locations all \atlas probes that violate the \sol.
Nearly half of violating probes claim to be in the USA (29\%), Germany (11\%), or France (9\%).
The majority of violating probes seem to respond from hundreds to thousands of kilometers from their operator-reported geolocation (Appendix~\ref{app:distribution_errors}).
%We find only one violating probe less than 10~km away from the geographic center of its residing country (Taiwan), suggesting that \atlas may have used a default geolocation for this probe.

\begin{figure}
    \centering
    \includegraphics[width=\columnwidth]{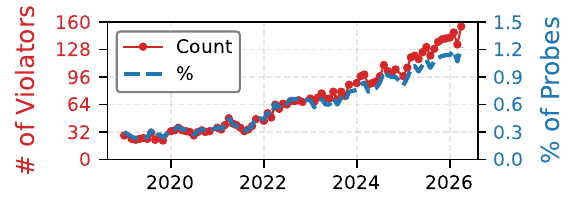}
    \caption{\textbf{Violations over time}---\textnormal{The number of violating probes increased from 22~probes in January 2019 to \numviolatingtoday~probes in \presentday. The relative proportion of these probes quintupled during this interval.}}
    \label{fig:violations_over_time}
\end{figure}
Across five~years, the raw number of probes that violate SOI increases every year.
Using historical data from built-in RIPE measurements, Figure~\ref{fig:violations_over_time} illustrates an increase in number and proportion of violating probes, from just 22 (0.23\%) in January~2019 to \numviolatingtoday (1.23\%) in \presentday.
The increasing trend is not driven by older probes accumulating repeated violations: violating probes are significantly younger than other probes.\footnote{The median probe age is 837 days for violators and 1503 days for non-violators (Mann–Whitney $p<10^{-200}$, Common-Language Effect Size = 0.38).}
In total, \totalallyears probes (2.4\% of all probes) seem to be initiating from a different geolocation at least one point in time.
Nearly half of probes (345/\totalallyears) eventually stop violating SOI, with 35\% of those eventually disconnected or abandoned after the SOI is violated.

\begin{figure}[t]
    \centering
\includegraphics[width=\linewidth]{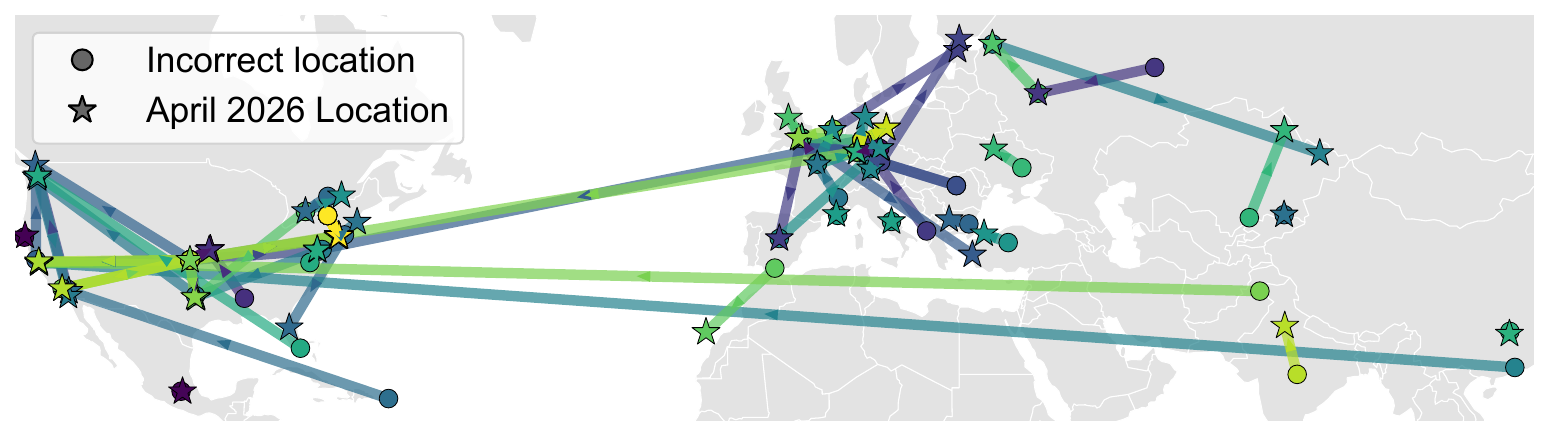}
    \caption{\textbf{Operators updated their location after our report}---\textnormal{Shown are all operator-updated probes above the equator, including substantial cross-country and transcontinental moves.}}
    \label{fig:updated_post_dispute}
\end{figure}

\subsection{Correcting Misreported Geolocation}
\label{sec:disputes}

We work with \atlas to correct misreported probe geolocation.
Between July~2025 and \presentday, we reported to \atlas 258 probes that were violating SOI constraints at the time of reporting.
\atlas then notified operators that their reported probe geolocation was inconsistent with network measurements. 
Operators could accept or deny the proposed correction, update the probe location when appropriate, and optionally provide a comment.

As of April~2026, 103 operators responded to our notifications: 93 accepted and updated their operator-reported geolocation, while 10 denied that their location was inaccurate. 
Notably, 4\% of probes whose locations were corrected had violated SOI on only one day, suggesting that even sparse violations can reveal real misreported geolocation. 
Figure~\ref{fig:updated_post_dispute} plots operator-corrected locations above the equator, where most violating probes are located, and shows multiple intercontinental updates.
This is consistent with our historical analysis: 50\% of operators whose probes no longer violate SOI updated their geolocation over 1,000~km, indicating that their probes could have incorrectly reported locations on different continents (Appendix Figure~\ref{fig:loc_ch}).

Some denied reports involve probes hosted on third-party infrastructure, rather than directly operated by the probe owner. Of the ten operators who denied that their location was wrong, four provided comments stating that their probe ran on a virtual private server hosted by a third-party provider. These operators reported that they had confirmed the leased server was in the country listed by the probe. We manually checked all 10 cases and found that they either do not appear to be country-level violations or are located near borders, but still appear largely inconsistent with the reported city-level location.
Interestingly, hosting-network probes---probes for which the operator themselves lack ground truth---are disproportionately likely to violate \sol. 
Using the March 2026 ASdb snapshot~\cite{ziv2021asdb}, we classify probes based on the ASNs announcing their IP addresses. Although most violating probes are ISP-hosted, ISP probes are significantly less likely to violate \sol overall ($p = 6.19 \times 10^{-4}$, risk ratio = 0.56), whereas probes in hosting networks are significantly more likely to do so ($p = 1.81 \times 10^{-6}$, risk ratio = 2.53).\footnote{We use a two-proportion z-test to assess statistical significance. Effect sizes are reported as risk ratios, defined as the ratio of the probability of violation within a given class (e.g., ISPs or hosting providers) to that of probes outside that class.}
%\enlargethispage{2\baselineskip}
\subsection{Understanding Contributing Factors} \label{sub:contributing}
Through a manual analysis of reported geolocations and round trip times, we discover that tardiness in updating geolocation after a move, and initial misconfiguration of geolocation are the most likely and common contributors to SOI violations.

\subsubsection{Tardiness} \label{subsub:tardy}
\begin{figure}[t]
    \centering
\includegraphics[width=\columnwidth]{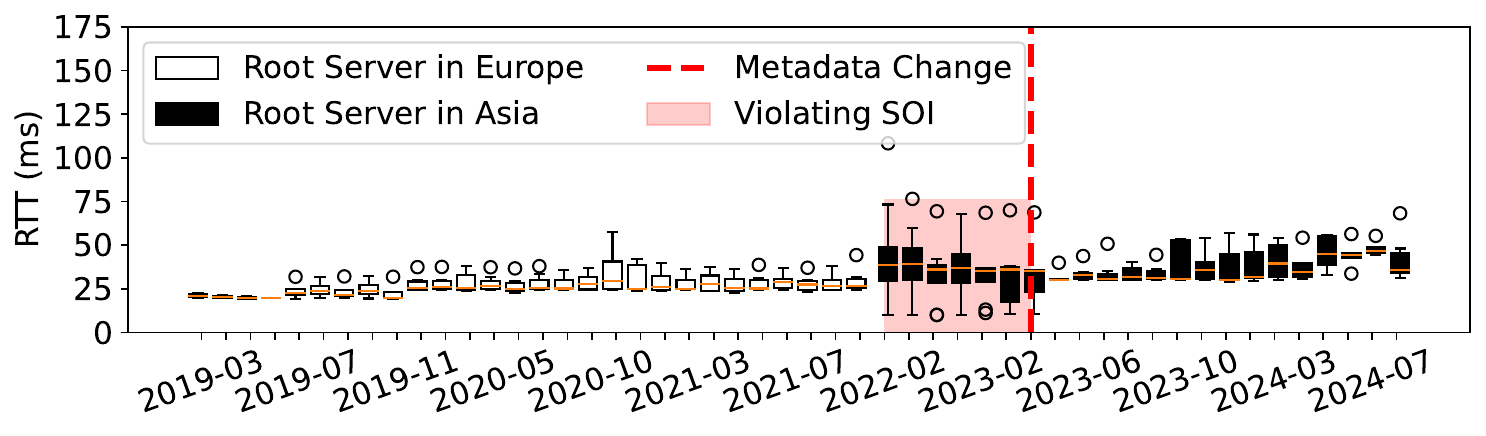}
    \caption{\textbf{Moving from Germany to Brunei Case Study}---\textnormal{The operator's probe reports a SOI violation between February 2022 and April 2023, before updating their location  from Germany to Brunei in April 2023.}}
    \label{fig:case_study}
\end{figure}
\vskip 0.2cm
At times, probes reporting RTTs that violate SOI are simply late to report a move.
We see 41\% of violating probes appear to stop violating SOI after the operator updates the probe location.
%\todo{Furthermore, 12.5\% of all violating probes only violated once in our data.}
For example, Figure~\ref{fig:case_study} shows longitudinal measurements from a probe whose operator reported a location in Germany.
From 2019 to 2022, its minimum RTTs to three root letter instances hosted in Germany were compatible with the operator’s claimed Germany location ($\approx$19~ms).

However, after a brief disconnect in 2022, the root server responses changed drastically.
Subsequent measurements show RTTs under 10~ms to two root letters in Bangkok, implying the probe should be within 1000~km of Bangkok, even though its geolocation still shows a location in Germany.
Finally, in 2023, the operator updated the probe’s geolocation to Brunei, which is consistent with the inferred proximity to Bangkok.
\looseness=-1
%After the metadata update, we no longer observe SOI-violating RTTs.

\subsubsection{Initial Misreporting.} \label{subsub:violator_at_birth}

Some long-term violations stem from probes configured with incorrect geolocation at deployment.
A quarter (23.3\%) of violating probes appear to violate SOI their entire existence (either disconnecting or still violating as of \presentday).
An additional 12.72\% appear to violate SOI from their initial misconfiguration until the operator updates the location (Section \ref{subsub:tardy}).

Consider Probe 1000011, which violated the \sol from its deployment in October 2019 until June 2023.
It reported a location in California until 2023, when the geolocation updated to Virginia. However, its minimum RTT across all 12 root letters remained stable---unexpected for a coast-to-coast move. 
Moreover, RTTs to root servers located along the United States East Coast (e.g., Miami, Florida; Dulles, Virginia; Washington, D.C.) violated \sol at the California location, but stopped once the location was updated. This suggests the probe was likely near Virginia all along.
The location change coincides with an upstream provider switch from Cogent to Lumen, hinting that a network change may have prompted the operator to verify the probe's configuration and update the geolocation. Roughly half (43/93) of responding operators in our RIPE campaign from Section~\ref{sec:disputes} appear to have initially misconfigured their probe location.

\subsubsection{Lack of Violation-Resolving Update.}
\label{sub:dependent}
The remaining 35\% of probes (256/\totalallyears) either never update their reported geolocation (187/256) or update with no violation resolution (69/256), even though their violation patterns change over time. 
For some probes, their geolocation remains unchanged through \presentday, despite abrupt changes in RTTs to the same root instances or to an entirely different set of instances (see Appendix Figure~\ref{fig:hypothesis}).
For others, violations are observed only with respect to a consistent subset of DNS instances; consequently, if those instances do not respond, no violation is observed.

% \subsubsection{Responding Instance Patterns.}
% \label{sub:dependent}
% An artifact of our methodology is that we cannot control which anycasted DNS instance responds on behalf of the root letter.
% Therefore we see probes with bursts of violating time periods intermittent with no violations.
% We calculate the set of DNS root hostnames that respond to each probe when it violates \sol and when it does not.
% 29\% of violating probes have a strict set of hostnames they violate to over time. 
% An additional 5\% of violating probes have a strict set of locations they violate to over time, regardless of root letter.
% Furthermore, a probe need only violate at least 2 independent root letters in one snapshot --- 4\% of the probes we reported which spurred operator action only violated in one snapshot.
% In Section~\ref{sec:disputes} we will discuss how our reports led to operators updating the metadata of 93 probes, 56 of which fall into this broader category of Responding Instance Patterns.

% \subsubsection{Hypothesizing remaining factors.}
% The remaining 13~probes keep the same coordinates from their configuration until \presentday, but they begin to violate SOI for the hosts they did not previously (therefore not satisfying Section~\ref{sub:dependent}).
% In Figure~\ref{fig:hypothesis} in the Appendix, we graph an example of such a probe.
% We hypothesize these probes have moved either once or multiple times, without ever updating their metadata or disconnecting instead.

\subsection{Impact of Misreported Geolocation}

Relying on a small number of incorrectly geolocated probes can subtly cascade into many more incorrect city-level geolocations.
We first revisit Du et al.'s evaluation of RIPE IPmap's single-radius engine, which infers city-level target locations from RIPE Atlas latency measurements~\cite{ipmapeval}.
Comparing the 2,000 probes used in their study with probes that violated \sol during their measurement period (September~2019--January~2020), we find that 33 probes appear to violate \sol. These probes risk incorrect city-level geolocation for 4.3\% of the study's 8,000 candidate IPs.
We then analyze the past year of \atlas measurements scheduled by the RIPE IPmap single-radius engine~\cite{ripe-atlas-api}. During this period, IPmap geolocated 98K candidate IPs using latency to the closest RIPE Atlas probe among 2.8K probes. Of these probes, 28 are likely mislocated, cascading to 1,125 distinct target IPs potentially geolocated to the wrong city.

\subsection{Methodological Sensitivity}
\label{sec:ablation}
\label{sec:limitations}

Our methodology is conservative by design, but SOI-based filtering remains sensitive to the measurement infrastructure used as vantage points. 
We evaluate our methodology's sensitivity to two parameters: a 100~km location buffer and requiring agreement from at least two independent root letters. 
On average, removing the buffer adds 26~violations per month, while removing the two-letter requirement adds 43~violations per month. 
The 100~km buffer is necessary: 9.5\% of responding root instances are reported at default IATA coordinates, which may not reflect their true physical locations. We therefore include a buffer to account for the likely discrepancy between the reported and actual instance locations.
The two-letter requirement is important in practice: under a one-letter rule, we found 53 probes near Santiago, Chile that violated only a single I-root instance listed in Concepción (netnod-cl-ccp), while nearby root instances did not violate \sol. 
Notably, in Section~\ref{sec:min_fp}, we found the same I-root instance violated our sanity check.
We reported this case to I-root, who preliminarily agreed that the instance is likely in Santiago, not Concepción; our two-letter rule therefore prevented us from falsely labeling these probes as violators.
\looseness=-1

A comparison with Darwich et al.~\cite{darwich2023replication} further shows that SOI-based filtering is sensitive to which vantage points are used to evaluate probes. 
Using Darwich et al.'s experiment window (2023-04-24 to 2023-05-03), we identify the same number of violating probes (96), but only 63~overlap. 
The difference largely stems from the vantage points: Darwich et al. use validated anchors, while we use root DNS servers. 
For probes identified only by Darwich et al., their anchors are on average 150~km closer than our root DNS servers; conversely, for probes identified only by our method, root DNS servers are on average 234~km closer than their anchors (see Appendix~\ref{app:darwich_details}). 
Nevertheless, using root DNS servers enables analyses without issuing new measurements (e.g., avoiding the millions of \atlas credits required by~\cite{darwich2023replication} for a single snapshot) and allows for retrospective studies.

Our methodology's conservative choices reduce false positives, but also limit what our method can detect. 
Our focus on \atlas may not represent all vantage points with operator-reported geolocations, although its scale underscores the importance of improving its accuracy. 
Our method likely underestimates both prevalence and distance-error magnitude because true violations may remain undetected when paths are slower than ideal, congestion-free links. 
It may also miss violations in regions where root DNS servers are far from probes or served by few operators, since longer paths loosen RTT bounds and our method requires agreement from at least two distinct root-server operators.
Future work should explore new methods to detect SOI violations, including providing tighter---but still accurate---bounds for detecting violations.

% \section{Limitations}\label{sec:limitations}
% Our work faces several limitations: 
% (1) the focus on \atlas does not represent all vantage points with operator-reported geolocations, although its scale as the largest community vantage-point collection underscores the importance of improving its accuracy;
% (2) the methodology likely underestimates both the prevalence of violating probes and the magnitude of their distance errors, since additional true violations may remain undetected when vantage points operate slower than ideal, congestion-free links;
% (3) violations may be missed in regions where root DNS servers are far from probes, since longer paths impose looser RTT bounds and leave more tolerance for misreported locations;
% and (4) violations may be missed in regions served by few root-server operators, since our method requires violations against at least two distinct operators.

\vskip 0.5cm
    
    %\section{Conclusion}

%at  \href{https://github.com/kizhikevich/violating_ripe_probes}{$\texttt{https://github.com/kizhikevich/violating\_ripe\_probes}$}, 
    \bibliographystyle{ACM-Reference-Format}
    \bibliography{refs}
    \clearpage
    \appendix
\section{Appendix}\label{app}

\subsection{Ethics}
 All RIPE data used in this analysis is publicly available at \url{https://atlas.ripe.net/api/v2/measurements/}. The RIPE location data is anonymized by a few blocks. This work does not raise any ethical issues.
% Note from the CFP that this section must include a statement about
% ethical issues; papers that do not include such a statement may be
% rejected.

\section{Prior Work}
\label{app:prior-work}
Given the importance of geolocation, relatively few studies have validated the accuracy of RIPE Atlas probe locations. 
Gharaibeh et al.~\cite{gharaibeh2017look} were the first to perform a basic ``sanity check'' in 2017.
They considered the reported locations untrustworthy if they had default country coordinates (typically geographic centers) or if two probes were connected to the same router, but reported locations more than 100km apart---thereby identifying 24 questionable probes. 
In 2020, Darwich et al.~\cite{darwich2023replication} applied a recursive method based on speed-of-Internet violations, filtering 9 of 723 anchors and 96 of 10K probes. Their approach compared RTTs exclusively among RIPE Atlas anchors and probes, iteratively removing nodes whose latencies violated physical feasibility constraints.
Both methods detect groups with violations, but cannot isolate the specific mislocated probe.
Further, no work provides an in-depth 
analysis of where violations occur or what causes them.

These prior works rely on the same vantage points, the RIPE Atlas probes themselves, to identify other potentially violating RIPE Atlas probes.
In contrast, we use a publicly available set of globally distributed nodes that are subject to far stricter operational and management requirements: the DNS root servers.

\subsection{Sources of Discrepancies with Darwich et al.~\cite{darwich2023replication}}
\label{app:darwich_details}

Differences between our method and Darwich et al.~\cite{darwich2023replication} arise from both vantage point selection and data availability. 
Darwich et al. propose a RIPE Atlas sanitization procedure that (1) performs anchor-to-anchor measurements, (2) iteratively removes anchors with SOI violations, and (3) tests probes against the validated anchors. 
We extract the 96 violating probes reported in their open-source repository and reproduce their time window (2023-04-24 to 2023-05-03), identifying 96 violating probes with 63 overlapping between the two methods. 
Of the probes reported by Darwich et al., \atlas contains the necessary geolocation and root DNS measurements for 79 probes, and their repository is missing data for 3 probes. 
Among the remaining probes, we find three cases where the anchor used in their probe-anchor measurements violates \sol under our analysis, rather than the probe itself.

The remaining discrepancies are explained by differences in distances between probes and their respective vantage points. 
For probes identified only by Darwich et al., the destination anchors are on average 150~km closer to the probes than the root DNS servers used in our analysis, reducing the likelihood of observing violations under our method. 
Conversely, for probes identified only by our method, root DNS servers (See Figure~\ref{fig:dns_positions}, which plots the geolocations of the instances of each root letter) are on average 234~km closer than the anchors used by Darwich et al., reducing the likelihood of observing violations under theirs. 
Together, these results show that SOI-based sanitization depends on both the placement and selection of measurement infrastructure.

\section{Distribution of Probe Distance Errors}
\label{app:distribution_errors}

The majority of violating probes appear to be hundreds to thousands of kilometers from their operator-reported geolocations.
For every <probe, vantage point> pair that violates the SOI constraint, we calculate the minimum error of distance by 
(1) selecting all VPs that violated SOI,
(2) multiplying the minimum observed RTT by SOI, to derive the furthest distance, $d$, a probe could be if it traveled at exactly SOI,
(3) calculate the theoretical distance, $d_{theory}$, between our VP and the operator-reported probe location,
(4) subtract $d$ from $d_{theory}$, recording the minimum distance error. 

We analyze the full distribution of distance errors: nearly 50\% exceed 1,000~km ( Figure~\ref{fig:cdf_error}).
The large distance errors suggest violating probes are likely on different continents, as we discuss in Section~\ref{subsub:tardy} and show in Section~\ref{sec:disputes}.
Figure~\ref{fig:loc_ch} further shows that 50\% of operators whose probes no longer violate SOI updated their geolocation over 1,000~km, indicating that their probes could have incorrectly reported locations on different \textit{continents}.

\begin{figure}[t]
    \centering
    \includegraphics[width=\columnwidth]{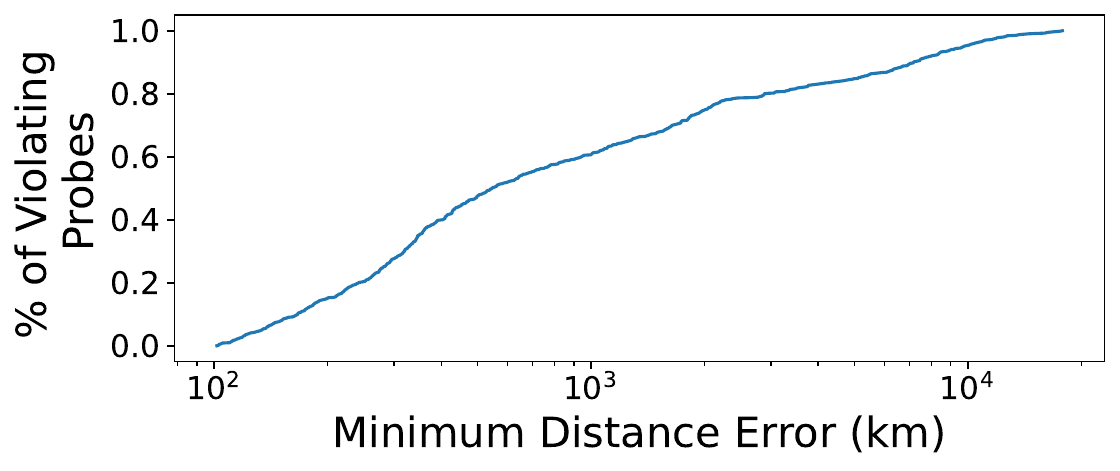}
    \caption{\textbf{Distance Error}---\textnormal{Nearly half of likely misreported probes are likely over 750~kilometers away from their reported locations.}}
    \label{fig:cdf_error}
\end{figure}

%% keep this figure! A reviewer asked for it!
\begin{figure}[h!]
    \includegraphics[width=\columnwidth]{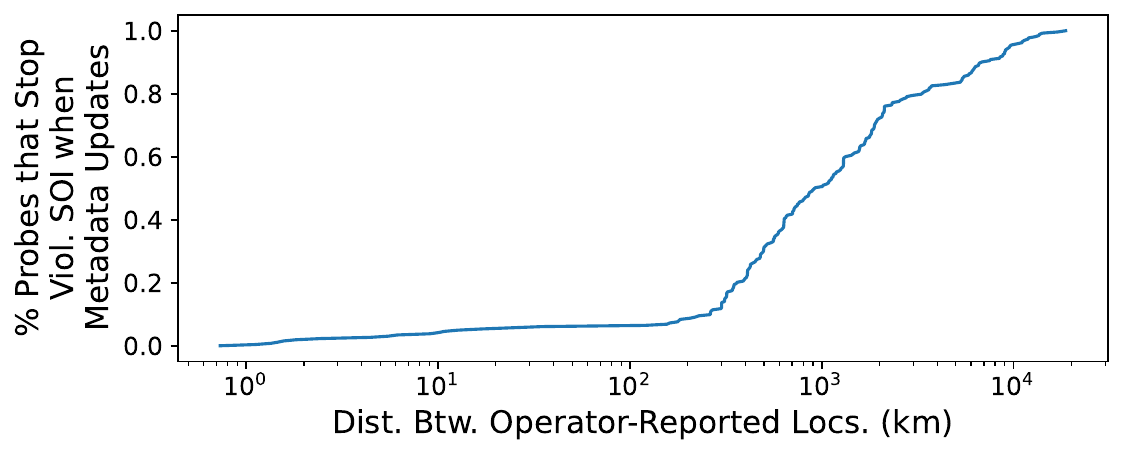}
    \caption{\textbf{Distance Between Reported Locations Pre- and Post-Violations}---\textnormal{Nearly half of operators resolved violations by updating locations over 1,000~km.}}
    \label{fig:loc_ch}
\end{figure}

\section{Time to Fix Geolocation}

The majority of violating probes take at least 13~weeks to update their geolocation.
Figure~\ref{fig:late} plots the time from when a probe violates the SOI until an operator updates its geolocation such that it no longer violates SOI. 

\begin{figure}[h!]
    \includegraphics[width=\columnwidth]{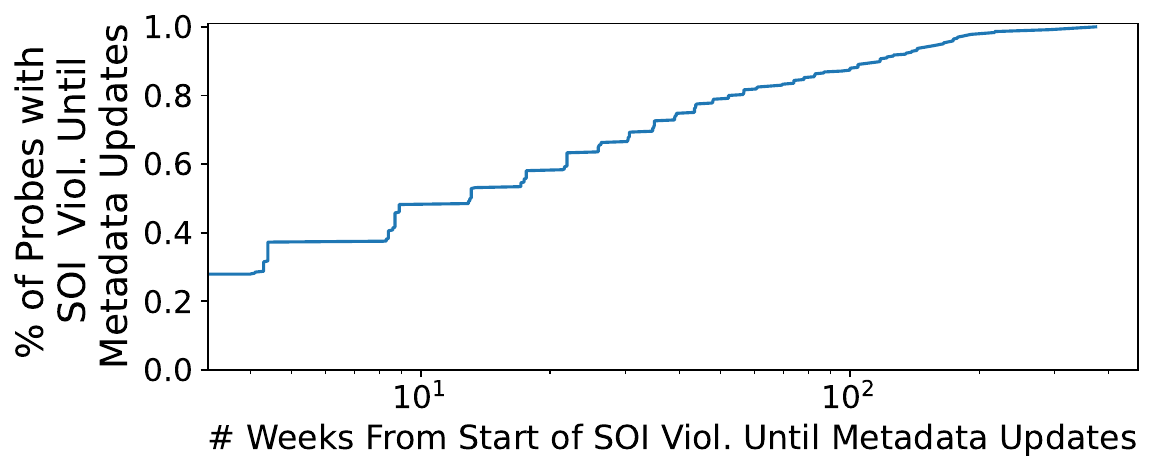}
    \caption{\textbf{Number of Weeks Between SOI Violation and Geolocation Update}---\textnormal{Of the probes that do not disconnect, it takes over 13~weeks for 50\% of such probes to update their geolocation. Notably, the SOI violation no longer occurs.}}
    \label{fig:late}
\end{figure}

% \subsection{Our vantage point coverage}
% In Figure~\ref{fig:ark_map}, we plot all vantage points we control. Notably, our vantage points are thoroughly represented across six continents. 

% \begin{figure}[h!]
%     \centering
%     \includegraphics[width=\columnwidth]{figs/world_map_ark.pdf}
%     \caption{\textbf{Vantage Point Coverage}---\textnormal{We plot the 294 globally distributed vantage points we use as ground truth.}} 
%     \label{fig:ark_map}
% \end{figure}

\section{Analysis Continued}
\begin{figure}
        \centering
        \includegraphics[width=\columnwidth]{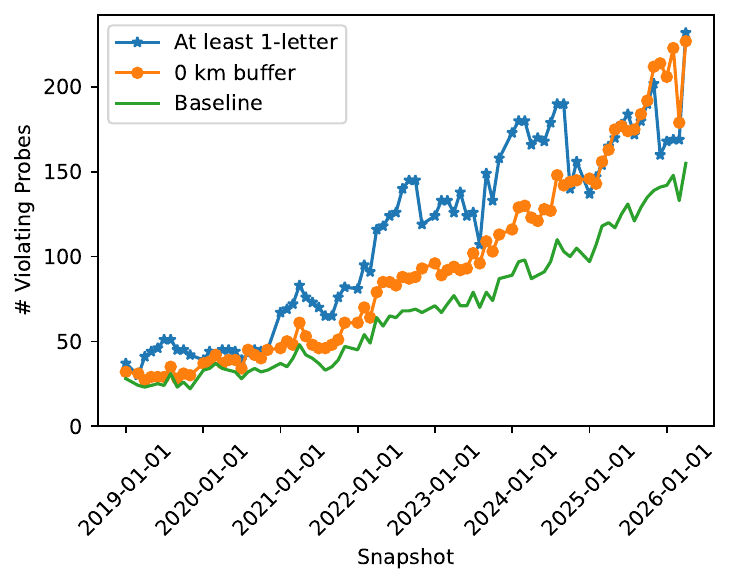}
        \caption{\textbf{Testing removal of 100~km buffer and independent root letter agreement}---\textnormal{On average, removing the buffer adds 26 violations per month, while removing the two-letter requirement adds 43 violations per month.}}
        \label{fig:ablation}
\end{figure}
In Figure~\ref{fig:ablation}, we show how the number of violating probes follows similar temporal patterns regardless of removing the 100~km buffer or multi-letter agreement criteria. 

\begin{figure}
        \centering
        \includegraphics[width=\columnwidth]{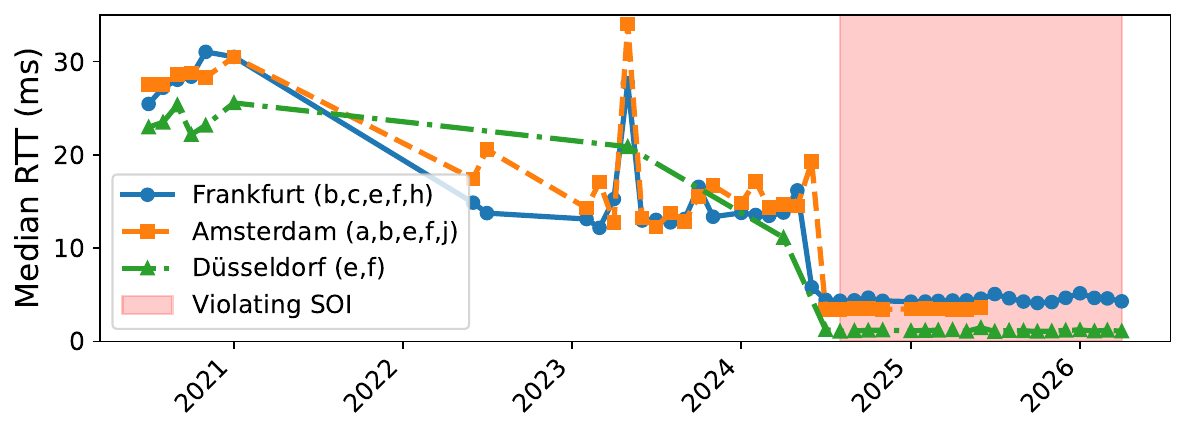}
        \caption{\textbf{RTTs of a probe that likely moved, but has yet to update its geolocation}---\textnormal{This probe shows a significant difference in RTTs across numerous root instances (see the letters in the legend) in the same cities over time.}}
        \label{fig:hypothesis}
\end{figure}

In Figure~\ref{fig:hypothesis}, we show how a probe could keep the same metadata over time, but experience vastly different RTTs to the same host sites over time. While we cannot say for certain the probe physically moved, the sudden drop in RTTs to Frankfurt, Amsterdam, and Düsseldorf would suggest the probe may have moved closer to these locations.

\begin{figure*}
        \centering
        \includegraphics[width=\linewidth]{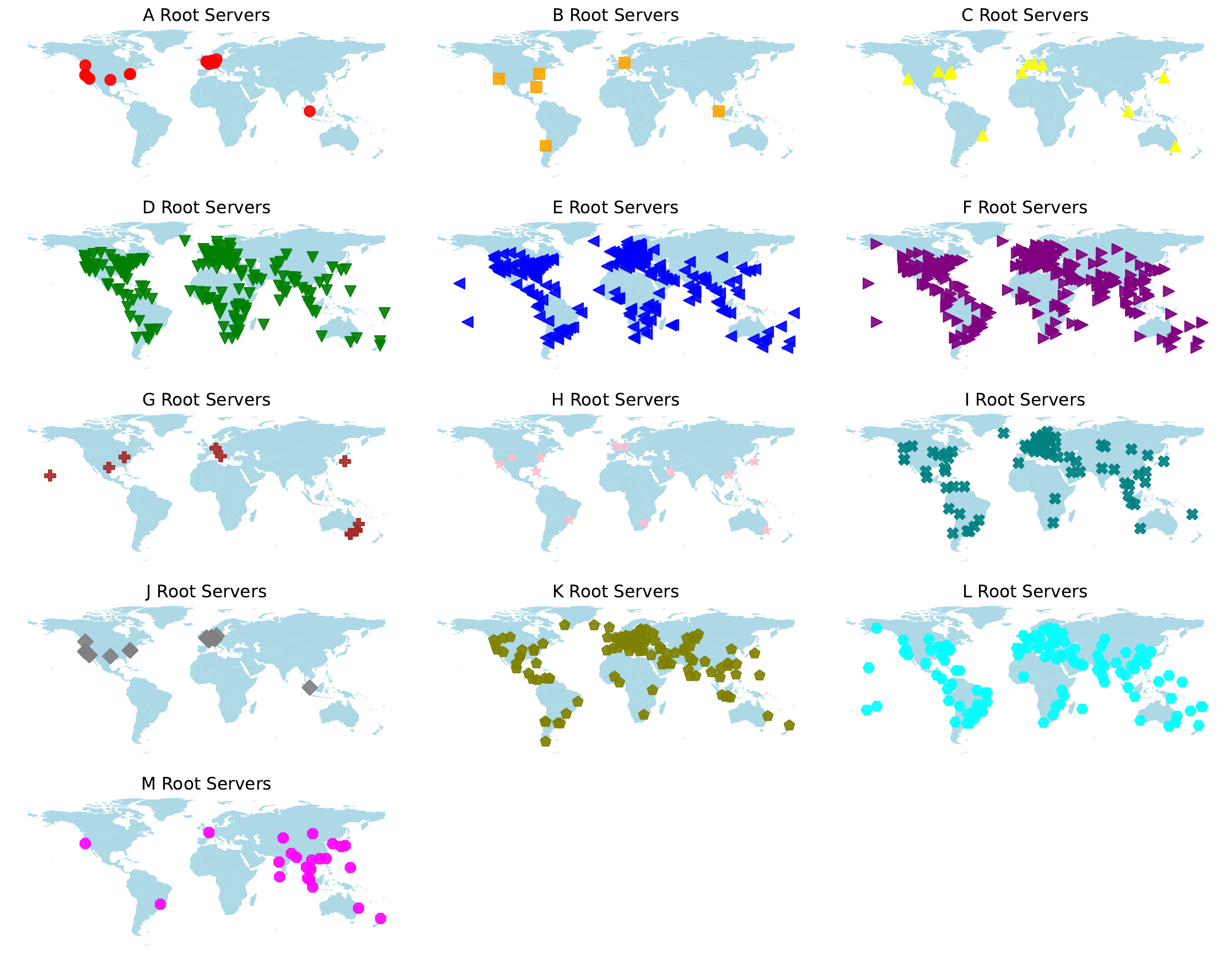}
        \caption{DNS Root Letter Instance Geographic Distribution.}
        \label{fig:dns_positions}
\end{figure*}

\fi

\ifimc
    \begin{abstract}
Geolocation plays a critical role in understanding the Internet. 
In this work, we analyze and fix operator-misreported geolocation.
Using DNS root servers to detect speed of Internet violations, we conservatively infer that at least 2\% of vantage points in the largest community-vantage point collection, RIPE Atlas, were not located at their operator-reported geolocation between January 2019 and \presentday. 
%Furthermore, the problem is escalating: within the past seven years, the number of probes likely reporting the wrong location has increased over five-fold. 
To increase the accuracy of future studies that use operator-reported geolocation, we open source our simple methodology, implement a reporting campaign of violating probes within RIPE, work with operators to fix the misreported geolocations, and release a continually updated dataset of RIPE vantage points that likely misreport geolocation.
\end{abstract}
    \maketitle

    \section{Conclusion}

Operator-reported geolocation remains a critical but weakly validated input to Internet measurement studies. 
We show that misreported probe locations can persist for months, appear thousands of kilometers from their reported locations, and cascade into downstream geolocation errors. 
Critically, we take a first step toward correcting misreported geolocation at the source by working with RIPE Atlas to report and fix physically impossible reported geolocations to operators. 
Further, we open-source our methodology and maintain a continually updated list of currently violating \atlas probes.
A more accurate foundation for Internet measurement begins with rooting out the physically impossible locations hidden in today's community-reported datasets.

\looseness=-1

    \bibliographystyle{ACM-Reference-Format}
    \bibliography{refs}
    \clearpage
    
\fi

\ifOLDimc
    \input{gameplan}
    \begin{abstract}
Geolocation plays a critical role in understanding the Internet. 
In this work, we analyze and fix operator-misreported geolocation.
Using DNS root servers to detect speed of Internet violations, we conservatively infer that at least 2\% of vantage points in the largest community-vantage point collection, RIPE Atlas, were not located at their operator-reported geolocation between January 2019 and \presentday. 
%Furthermore, the problem is escalating: within the past seven years, the number of probes likely reporting the wrong location has increased over five-fold. 
To increase the accuracy of future studies that use operator-reported geolocation, we open source our simple methodology, implement a reporting campaign of violating probes within RIPE, work with operators to fix the misreported geolocations, and release a continually updated dataset of RIPE vantage points that likely misreport geolocation.
\end{abstract}
    \maketitle
    \section{Introduction}

IP address geolocation plays a central role in networking and security: 
understanding the Internet's topology to improve last mile latency~\cite{lastmilelatency, lastmilecongestion}, filtering traffic to prevent cyber attacks~\cite{wan20origins}, and identifying malicious activity~\cite{izhikevich2024using,dnsintercept}, all rely on geolocation. 
To geolocate an IP address, methodologies often rely upon a global, human-reported set of vantage points (i.e., probes) to measure from and verify accuracy against. 
For example, geolocation based on latency measurements often require a network of probes, each with operator-attested geolocation, to be located near every target to accurately estimate every target's geolocation~\cite{hu2012towards,gueye2004constraint,katz2006towards,wong2007octant}.

%Thus, such a set of vantage points need be geographically distributed, relying on operators to attest to the location of each probe.
% Yet, despite geolocation's importance for nearly two decades, we still have no method to geolocate IPs with 100\% accuracy~\cite{darwich2023replication}.

%latency is often distorted due to a diversity of routing paths, network types (e.g., optical fiber, copper wire, satellite, mobile), and congestion. 
% Geolocating IPs is non-trivial, in part because it requires a global set of vantage points.
% \changed{For example, geolocation methods that rely on accurate latency measurements require that a vantage point exist nearby every target~\cite{hu2012towards,gueye2004constraint,katz2006towards,wong2007octant}.}
% Verifying the global accuracy of a geolocation methodology requires that a large-enough set of vantage points with already-known geolocations exists to compare against.
%Verifying geolocation's global accuracy requires a large-enough set of distributed IPs with already-known geolocations to compare against (i.e.,  a ground truth dataset). 

While necessary, relying on human operators to report and maintain accurate probe geolocation can introduce human-error to the geolocation process.
For example, if an operator moves their probe, but fails to update its location metadata, geolocation methodologies relying on the outdated information risk inaccurate conclusions.
To account for incorrect metadata, a few geolocation studies take an additional precaution to filter out probes reporting unlikely locations~\cite{darwich2023replication, gharaibeh2017look}.
%Since filtering erronous vantage points is not the broader intention of geolocation studies,
Unfortunately, existing filtering methodologies often require high bandwidth~\cite{darwich2023replication} and risk filtering accurately geolocated probes~\cite{gharaibeh2017look,darwich2023replication}.
Furthermore, probes with misreported geolocations are still not well understood, including where they are most likely to be located, what their magnitude of error is, and how likely they are to eventually be corrected. 
As operator-reported geolocations continue to serve as a foundation for many studies~\cite{bajpai2015lessons,holterbach2015quantifying,iodice2019periodic,fanou2015diversity,corneo2021surrounded,gigis2017characterizing,sukhov2014evaluating,davisson2021reassessing,candela2019using}, it is critical that we understand the successes and pitfalls of our community-contributed datasets and work towards an accurate ground truth.

%: many questions remain about the incorrectly-reported vantage points: \TK \TK 
%Mainly, it is unclear whether such a filtering introduces geographic biases against certain regions, and what the magnitude of error of incorrect reportings is (i.e., a few miles of error is not nearly as troublesome as a thousand).
% However, the geographic bias and magnitude of error of these incorrectly geolocated probes is still unknown.
% Likewise, the longevity of outdated probe metadata is also unclear, potentially impacting multiple studies.}

%ground 1st sent more in community vantage points
% \changed{ 
% However, the geographic bias and magnitude of error of incorrectly geolocated probes is still unknown.
% Furthermore, it is unclear whether the misreported location of these probes is ephemeral or a more persistent problem. 
% Without knowing the answer to these questions, it is unclear whether there are regions now lacking coverage and whether the prevalence of this problem is stagnant or worsening.}
% \liz{i think this is where we want to hint that some work has hinted that its not all accurate, but its not clear how prevelant the problem is--or whatever geoff's recommended framing was}

%, or are highly improbable.  

In this work, we provide an in-depth analysis of operator misreported geolocation.
We apply a simple methodology that conservatively infers the set of probes that do not respond from their operator-reported geolocation.
The methodology (1) uses speed of Internet calculations to derive a minimum bound of expected latency between two vantage points---the target vantage point and a vantage point we control---and 
(2) applies the minimum bound to infer whether operator-reported geolocations are physically possible. 

Between May~2024 and March~2025, we deploy our methodology from 294~globally distributed probes, to measure the validity of operator-reported geolocation of the largest and most geographically distributed community-vantage point collection: RIPE Atlas~\cite{ripeatlas}.
Our methodology infers a lower bound of 470 probes (3.96\%) likely not responding from their reported location between May 2024 and March 2025, as doing so would require exceeding the physical limits of Internet speed (e.g., the speed of light in fiber).
Our work identifies over 4~times the amount of violating probes, while applying a modified methodology that uses 9~times less iterative measurements (evaluated in Section~\ref{sub:sec:eval}) than prior work~\cite{darwich2023replication,gharaibeh2017look}.
%Compared to prior work~\cite{darwich2023replication,gharaibeh2017look}, our methodology identifies over twice the number of violating vantage points while using 9~times less probing packets.
% and 159 (1.33\%) present-day ``violating'' probes:

% In May~2024, we deploy our methodology from Ark~\cite{caidark}, 170~globally distributed probes we control, to measure the validity of operator-reported geolocation of the largest and most geographically distributed community-vantage point collection: RIPE Atlas~\cite{ripeatlas}.
% Our methodology discovers 197 (1.5\%) present-day ``violating'' probes: probes that are likely not responding from the operator-reported geolocation, because otherwise they would surpass the speed of Internet (e.g., speed of light in fiber). 
% Compared to prior work~\cite{darwich2023replication,gharaibeh2017look}, our methodology identifies over 2~times the amount of violating vantage points while using 9~times less bandwidth.

%In this work, we challenge the assumption that commonly used community-contributed vantage points accurately report geolocation
%Vantage points do not always respond from their operator-reported geolocation. 
%198/12.906 = 1.5
%393/ 37,918 = 1
%, or are highly improbable.  

%BIAS - some regions lose significant coverage and  hundreds of miles away
%We find that there is a geographic bias for violating probes, since not only is SOI violated, but the magnitude of the violation suggests the probe must be in an entirely different country or state.

Errors in misreported geolocation are substantial; roughly 50\% of violating probes likely responded at least 1,600~kilometers away from their reported location. 
Inaccurate geolocation disproportionately affects countries with low vantage point coverage, such as southern Africa.
Therefore, entire countries (e.g., Lesotho, Botswana) previously reported to have at least one vantage point are left with no verified \atlas coverage, potentially impacting the accuracy of the prior work that depends upon them~\cite{surr_by_clouds,NREN_africa,fanou2017four}. 

%50% of operators take more than 7 weeks to update and it's only getting worse, increasing evry year
We consider how inaccuracies prevail over time, finding that operators are slow to update metadata and that the number of violating probes is increasing. The majority (52\%) of geolocations are updated after the vantage point violates the speed of Internet for at least 7 weeks, suggesting most violating vantage points misreport locations for months. 
Furthermore, the number of violating probes increases every year, highlighting that this problem is not diminishing.

%We propose a new automated methodology for researchers to filter out vantage points that are responding from improbable geolocations. 

Today, many studies do not report filtering for violating probes.
We encourage future studies relying on operator-reported geolocation (e.g., geolocation studies) to use our open-source methodology to filter for a more-validated set of probes. 
%We show that this filtering step is simple, but often necessary, to avoid exposing the ground truth to outliers that are thousands of kilometers away from their reported location. 
%---using only open-source data and no measurement credits, we are able to find 196 globally distributed RIPE probes that violate the speed of Internet today, in only 18 minutes. 
%We encourage future studies seeking accurately geolocated vantage points to run our open-source methodology to achieve a more validated set of distributed probes. 
To further lower the barrier to validate the dataset that numerous studies rely upon~\cite{surr_by_clouds,NREN_africa,fanou2017four,darwich2023replication,dnsintercept, lastmilecongestion, lastmilelatency}, we release a list of 664 likely-inaccurately-geolocated RIPE Atlas probes from the past five~years and maintain a weekly-updated list of currently-violating probes at [anonymized].

% As our community moves forward with improving geolocation, it behooves us to ensure that the ground truth we are using is correct. 
% Thus, we automate and open source our methodology to help researchers filter vantage points that are responding from improbable geolocations.
% We implore the community to validate prior work that relied upon the inaccurate geolocation and validate self-reported geolocation for all future studies. 
% We also release a list of the likely-inaccurately-geolocated vantage points and are working with RIPE Atlas to resolve these issues.

%results - topic sentence refers to the brosder RQs
%meth: 1.5% of probes violate SOI
%how innaccurate: hundreds of miles away
%some regions lose significant coverage
%50% of operators take more than 7 weeks to update
%it's only getting worse, increasing evry year
    \section{Background and Related Work}
Many studies rely on the RIPE Atlas platform to develop new geolocation methodologies and, more broadly, understand the Internet. These studies are inherently dependent on the accuracy of operator-reported geolocations, motivating us to find vantage points with misreported geolocations, more deeply understand the inaccuracies, and lower the barrier for researchers to validate probes.

\vspace{3pt}
\noindent
\textbf{RIPE Atlas.}\quad
RIPE Atlas is a measurement platform consisting of tens of thousands of networked devices (i.e., vantage points) distributed world-wide.
%Each RIPE Atlas device is capable of a variety of measurement tasks such as ping/traceroute to a target IP address or domain, performing DNS resolution, and sending HTTP requests. \todo{sentence about built-in versus do it yourself measurement} 
As of March~2025, RIPE Atlas has 11,908 active vantage points across 222 countries and 8,356 ASes \cite{ripeatlas}.
RIPE Atlas vantage points include two types: anchors (hosted by organizations) and probes (hosted by volunteers).
Anchors are considered more ``accurate''~\cite{davisson2021reassessing} because hosts must purchase RIPE-approved hardware and either build the anchor themselves or install it on a high-performance virtual machine.
We refer to both anchors and probes as \atlas probes unless stated otherwise.

% Anchors are hosted, usually by organizations, in well-connected networks such as IXPs and data centers. 
% The hosts of RIPE Atlas anchors need to purchase hardware approved by the RIPE NCC and build the physical anchor themselves, or install the anchor software on a virtual machine that meet the performance requirements. 
% Probes are less powerful devices designed to be hosted in home or office networks and can be hosted by anyone willing to volunteer. Hardware probes are distributed by the RIPE NCC and volunteers also have the option to host their own software probe.
 
The RIPE NCC provides a convenient interface for participants to manage their probes: after the probe is connected, the probe host can see their probe metadata (e.g. prefix and AS number) on their RIPE Atlas dashboard and report the physical location of their probe by placing a pin on a map. The reported probe location will be skewed at most by 1 kilometer to protect the probe host's privacy~\footnote{We observed the same anonymization for the 6 \atlas probes we control.}. When a probe becomes disconnected, the RIPE Atlas platform will regularly send the host email notifications to remind them of the disconnection. 
Critically, if a host physically moves the probe, they must manually update their reported location. 
Should a probe's metadata location be incorrect, the RIPE platform allows users to dispute operator-reported geolocation. As of March~2025, only 3~probe locations are disputed.

\vspace{3pt}
\noindent
\textbf{Research Using RIPE Atlas.}\quad
\label{sub:sec:research_related}
The distributed architecture of the \atlas platform has proven useful in many research areas, especially in Internet infrastructure geolocation~\cite{dnsintercept, lastmilecongestion, lastmilelatency,surr_by_clouds,NREN_africa,fanou2017four}. 
Scheitle et al.~\cite{hloc} used \atlas probes to verify geolocation inferred from location hints embedded in reverse DNS records. 
Weinberg et al.~\cite{proxylie} used \atlas probes to measure the latency to 2K proxy and VPN providers, and found at least 1/3 of VPNs falsified their advertised geolocation.
Darwich et al.~\cite{darwich2023replication} used RIPE Atlas to replicate the results of previous geolocation works. 
Gharaibeh et al.~\cite{gharaibeh2017look} used \atlas probes to validate four geolocation databases. 
%probes and anchors as vantage points and targets 
RIPE themselves use \atlas probes to power a public IP geolocation tool, IPmap~\cite{ripeipmap,ipmapeval}.

% Given their known locations, RIPE Atlas probes can serve as vantage points capable of sending traffic to target IP addresses. The RIPE NCC has created IPmap, a public IP geolocation tool \cite{ripeipmap}. 
% One of its components, the IPmap single-radius engine, uses RIPE Atlas probes around the world to ping the target IP address and estimates the geolocation of the target based on the lowest obtained latency \cite{ipmapeval}. 

%In addition, RIPE Atlas can also be used as measurement targets to evaluate the accuracy of geolocation techniques.  %but were not able to achieve the same level of accuracy or coverage. 

%NOT ALL METHODS USE LATENCY-BASED METRICS!! please keep "often" or "most" do not change!

Many geolocation studies using RIPE Atlas probes---whether as vantage points or targets---rely on latency and are sensitive to location accuracy. Misreported probe locations can lead to inconclusive or incorrect results. For instance, if RIPE IPmap sees a 0.2~ms RTT to a mislocated probe, it may wrongly assign the target IP to that incorrect location.
Probes with incorrect reported locations have a larger impact on results where RIPE Atlas probes are sparsely deployed.
Candela et al. \cite{atlasforgeo} studied the maximum theoretically achievable geolocation accuracy for RIPE Atlas-based techniques, and found that regions with limited probes, such as Africa and Asia, have lower maximum accuracy. 
Trammell and Kühlewind~\cite{luckgeo} also found that high accuracy relies upon selecting a vantage point close to the target. 
%Such luck would be defeated in a situation where a probe reporting an incorrect location is selected.

Few studies have validated the accuracy of RIPE Atlas probe locations. 
Gharaibeh et al.~\cite{gharaibeh2017look} were the first to perform a basic ``sanity check'' in 2017.
They considered the reported locations untrustworthy if they had default country coordinates (typically geographic centers) or if two probes were connected to the same router, but reported locations more than 100km apart---thereby identifying 24 questionable probes. 
In 2020, Darwich et al.~\cite{darwich2023replication} applied a recursive method (see Section~\ref{sub:sub:sec:comp_prior}) based on speed-of-Internet violations, filtering 9 of 723 anchors and 96 of 10K probes. 
Both methods detect groups with violations, but cannot isolate the specific mislocated probe, risking the exclusion of accurate ones.

%original longer version
% To the best of our knowledge, few works validate the correctness of RIPE Atlas probe locations. 
% Gharaibeh et al.~\cite{gharaibeh2017look} in 2017 were the first to include a preliminary ``sanity check'' of RIPE Atlas probe locations. 
% They considered the reported locations untrustworthy if they had default country coordinates (typically geographic centers) or if two probes were connected to the same router, but reported locations more than 100km apart. 
% They identified 24 probes with potentially incorrect geolocation. 
% In 2020, Darwich et al.~\cite{darwich2023replication} introduced a recursive-like methodology, described in Section~\ref{sub:sub:sec:comp_prior}, to identify probes that violated the speed-of-Internet constraint, filtering 9/723~anchors and 96/10K probes.
% Notably, both methodologies identify groups of probes that contain at least one geolocation violation, but they do not pinpoint which specific probe within the group is responsible.
% Thus, both methodologies risk filtering probes that accurately report their geolocation. 

%keep the darwich and gharaibeh sentences in the same paragraph, they have a topic sentence tying them together
%keep how our work is different a sep. paragraph

Our approach identifies 4~times more violating probes while using 9~times less bandwidth (see Section~\ref{sub:sec:eval}) than prior work. 
We also provide the first analysis of violator locations, causes of misreporting, and impact on ground truth.
\looseness=-1
%liz: included in into
%\todo{1. SHould I include research using Atlas that are not sensitive to location correctness, like topology studies -- which ISPs have the most last mile latency\cite{lastmilelatency, lastmilecongestion}, and which ISPs do DNS interception \cite{dnsintercept}}

%liz: related reads well! motivation is good, we will keep related work as second section
%\todo{2. This also kinda feels like a intro/motivation...}
    \section{Methodology}
\label{sec:meth}
Our objective is to identify whether a probe responding from its operator-reported geolocation is physically impossible.
%, identifying a minimum bound of the number of inaccurately geolocated probes.
We rely on simple physics to infer whether measured latency between our measurement source and the destination's reported location exceeds the speed at which light travels in either a fiber optic cable or an inter-satellite laser. 
We measure latency using a network of probes operated by collaborators (\ark\footnote{Anonymized for submission}) and historical \atlas\ measurement data.

\vspace{3pt}
\noindent
\textbf{Vantage Points.}\quad
Our methodology requires measuring latency to and from vantage points with validated locations. 
First, we use \ark, a set of 294~globally distributed vantage points whose exact latitude and longitude we verify directly with the operators.
\ark nodes are in 194~autonomous systems, across 218~cities, in 68~countries, on 6~continents (see Figure~\ref{fig:ark_map} in the Appendix for a map of coverage). 
%Notably, Ark is thoroughly represented in Western and Central Europe, as well as most regions of the United States.
\ark is open to researchers upon request.
% We plot the geographic distribution of \ark in Figure~\ref{fig:ark_map} in the Appendix.
% We deploy \ark, 
Additionally, we use seven RIPE operated ``central'' servers, located in 
%built-in ping measurements\footnote{\atlas collects ping data from every probe and anchor to each of these servers every 240 seconds.} from the past 5 years to the seven, stationary RIPE central servers: 
Fremont, California, USA; Newark, New Jersey, USA; Singapore; two in Amsterdam, Netherlands; and two in Nuremberg, Germany. 

\vspace{3pt}
\noindent
\textbf{Latency Measurement.}\quad
First, we measure the latency from all of our vantage points (VPs) to all 12K \atlas probes. 
We use the \ark nodes to ICMP ping every RIPE probe twice a month from May~2024 to March~2025.
Since RIPE Atlas is volunteer-run and intended for network measurements, ICMP filtering is unlikely to occur.
Between every <VP, probe> pair, we save the minimum round trip time (RTT) of all pings as the representative RTT because the speed of light in optical fiber or air imposes a hard lower bound. Unlike average RTTs, which can be inflated due to queuing or congestion, the minimum RTT reflects the fastest observed path and cannot be artificially reduced.
For RIPE central servers, we use their API to collect historical ping data from each \atlas probe to each server until September 2024, when RIPE stopped running measurements to their central servers.
We collect the minimum RTTs three times a day (6h,12h,18h UTC), on one day a week, over the past five years of data.
% , and save the minimum RTT across all measurements. 
% Thus, we derive one latency measurement for every source and destination (\atlas probes) pair.
Continuously unresponsive probes are filtered out.
\looseness=-1
%\footnote{If a probe is continuously unresponsive, we do not consider it as a violating probe or an active probe.}

\vspace{3pt}
\noindent
\textbf{Estimating Distance Between Source and Destination.}\quad
Second, to help derive the theoretical minimum latency between a vantage point and probe, we estimate the minimum (direct) distance. 
We calculate the direct distance between the reported coordinates of every \atlas probe to every RIPE central server. 
RIPE does not disclose the exact locations of the seven central servers, so we calculate a maximum error radius
around each city center.\footnote{In Section~\ref{res}, we discover that speed of Internet violations were large enough such that the use of radius calculations did not change the outcome of our results.}
%such that we account for servers that may be on the outskirt of the city.
We subtract the error radius from the direct distance. 
We also calculate the direct distance between the locations reported by the probes on the day we send pings to the location of our \ark vantage points. 
Since we verify the exact location of the \ark vantage points with our collaborators, we do not need an error radius.

\vspace{3pt}
\noindent
\textbf{Deriving Speed of Internet Theoretical RTTs.}\quad
Once we estimate the minimum distance between each probe and server, we convert it to a minimum theoretical RTT. 
To the best of our knowledge, Starlink satellites are the theoretically fastest networking infrastructure due to their use of inter-satellite lasers, which operate at the speed of light in a vacuum~\cite{chaudhry2022optical}.
To find probes hosted over Starlink, we look for probes whose IPv4 or IPv6 addresses are announced by Starlink's autonomous system (ASN\,14593)~\cite{hitchhiking}.
For Starlink-hosted probes, we conservatively estimate the theoretical \textit{absolute minimum} RTT as $direct\_distance/c$, where $c$ is the speed of light in a vacuum, as an extra precaution for being conservative and estimating the lower bound.
%\footnote{Note that if a Starlink probe is mobile (e.g., on a ship~\cite{hitchhiking}) and thus has no singular location, such a probe is not fit for a platform that assumes stationary locations. At any rate, if the probe violates the speed of Internet, we flag the probe as a violation.}
Otherwise, we calculate the theoretical \textit{absolute minimum} RTT if the pings were sent at the speed of light in optical fiber, $direct\_distance/\frac{2}{3}c$~\cite{chaudhry2022optical}. In the remainder of the text we refer to the speed of Internet as SOI.

\vspace{3pt}
\noindent
\textbf{Identifying Violating Probes.}\quad
We identify an \atlas probe as ``violating'' (i.e., we believe it does not respond from its operator-reported geolocation) if any \atlas probe measured latency is lower than the theoretical minimum latency from the self-reported geolocation to any of the seven RIPE central servers or any of the 294 \ark servers. 
For RIPE central server historical data, we take note of when within the past 5~years the RTT violation happened, if not presently occurring. 
Furthermore, since RIPE halted their built-in measurements to their central servers, we continuously check whether the last set of violating probes from September 2024 have updated their location or disconnected from the platform. If not, we label them as still violating.

\vspace{3pt}
\noindent
\textbf{Precautions to Minimize False Positives.}\quad
We take precautions to avoid overclaiming and false positives, at the expense of increasing false negatives.
For example, it is possible that a probe's operator-reported geolocation is accurate, but a middlebox  located away from the operator-reported geolocation might be responding on the probe's behalf. To assess the extent of this issue, we run ping measurements from present-day violating \atlas probes to \ark servers and find that 67/159 probes still violated the SOI constraint. These 67 probes should be excluded from use as either vantage points or measurement targets. The remaining 92 probes, while not suitable as measurement targets, may still be acceptable for use as vantage points.
Thus such probes could still be fit to measure from, but are best not to measure \textit{to}.
%Thus, we only make claims about the geolocation of where the probe responds from and not the probe itself.
% \changed{We examined the tags of the violating probes and found 23 with \texttt{NAT} tags. We manually examined traceroute measurements originated from those Atlas probes and observed that all probes had < 0.8ms RTTs to their gateways. Those RTTs were smaller than the last-mile latencies reported by~\cite{lastmilelatency} and thus we do not consider reported-NATs to impact the accuracy of our results.} 

%We avoid the use of statistical tests to determine the likelihood of a geolocation violation, because we seek to guarantee lower bounds, not a probability of lower bounds.
%In Appendix~\ref{app:stats}, we show how statistical tests and other traceroute-based methodologies lead to inevitable false positives.
%Thus, we only use strict speed of Internet bounds to determine geolocation violations.

We never attempt to identify the true geolocation of a probe, which is an open problem~\cite{darwich2023replication}. 
We only identify when the probe's response from an operator-reported geolocation is physically impossible. 
We discuss the implications and limitations of our claims in Section~\ref{sec:limitations}.
\vskip 0.5cm

    \section{Results}
\label{res}
Between May~2024 and March~2025, we conservatively infer that at least 470~unique probes violate \sol.
Of these, we primarily analyze the 159~probes that remain in violation as of March 2025.
These probes are geographically widespread, often responding from hundreds of kilometers away from their operator-reported geolocation, and leave entire countries without the majority of their original probe coverage.
The majority (76\%) of violating probes are likely due to operators misreporting geolocation.
Slow operator response times exacerbates the issue: 50\% of probes violate for over 7~weeks until operators update their geolocation.

\begin{table}[t]
\small
\centering
\begin{tabular}{l|r|r}
\hline
\textbf{Country} & \textbf{\# Violations} & \textbf{\% VPs} \\
\toprule
Germany & 51 & 2.69\% \\ %\ltgrey
\rowcolor{gray!10}
USA & 27 & 1.48\% \\
Netherlands & 9 & 1.32\% \\ %\ltgrey
\rowcolor{gray!10}
France & 8 & 0.75\% \\ %\ltgrey
Russia & 7 & 1.34\% \\
\rowcolor{gray!10}
Great Britain & 5 & 0.84\% \\
Czech Republic & 4 & 1.24\% \\ %\ltgrey
\rowcolor{gray!10}
South Africa & 3 & 2.38\% \\
Finland & 3 & 2.00\% \\ %\ltgrey
\rowcolor{gray!10}
UAE & 3 & 11.11\% \\
\bottomrule
\end{tabular}
\caption{Top Reported Locations With Most Violating Probes---\textnormal{Violating probes (VPs) are most likely to occur in countries with a large number of RIPE Atlas probes, including Germany and the USA, as of March~2025.}}
\label{tab:top_raw}
\end{table}
\vspace{-5pt}

% \begin{table}[t]
% \small
% \centering
% \begin{tabular}{l|r|r}
% \hline
% \textbf{Country} & \textbf{\# Violators} & \textbf{\% VPs} \\
% %& & \textbf{\# of Probes} \\
% \toprule
% USA & 43 & 2.35\% \\ %\ltgrey
% \rowcolor{gray!10}
% Germany & 42 & 2.28\% \\
% France & 12 & 1.10\% \\ %\ltgrey
% \rowcolor{gray!10}
% Russia & 9 & 1.59\% \\
% Great Britain & 7 & 1.08\% \\ %\ltgrey
% \rowcolor{gray!10}
% Canada & 6 & 1.88\% \\ %\ltgrey
% Netherlands & 6 & 0.95\% \\
% \rowcolor{gray!10}
% Spain & 5 & 2.19\% \\
% South Africa & 5 & 4.76\% \\ %\ltgrey
% \rowcolor{gray!10}
% India & 4 & 2.30\% \\ %\ltgrey

% \bottomrule
% \end{tabular}
% \caption{Top Reported Locations With Most Violating Probes---\textnormal{Violating probes (VPs) are most likely to occur in countries (as reported in the metadata) with a large number of RIPE Atlas probes, including the USA, Germany, and France, as of May~2024.}}
% \label{tab:top_raw}
% \end{table}
% \vspace{-5pt}
\begin{table}[t]
\small
\centering
\begin{tabular}{l|r|r}
\hline
\textbf{Country} & \textbf{\# Violations} & \textbf{\% VPs} \\
\toprule
Lesotho & 1 & 100.00\% \\ %\ltgrey
\rowcolor{gray!10}
Botswana & 1 & 100.00\% \\
Mozambique & 1 & 33.33\% \\ %\ltgrey
\rowcolor{gray!10}
Zambia & 1 & 25.00\% \\
UAE & 3 & 11.11\% \\ %\ltgrey
\rowcolor{gray!10}
Malaysia & 2 & 5.88\% \\
Albania & 1 & 5.00\% \\ %\ltgrey
\rowcolor{gray!10}
Moldova & 1 & 4.17\% \\
Argentina & 1 & 3.03\% \\ %\ltgrey
\rowcolor{gray!10}
Mexico & 1 & 2.70\% \\
\bottomrule
\end{tabular}
\caption{Geographic Skew of Misreported Probes---\textnormal{Countries with the largest fraction of misreported probes in March~2025 are near southern Africa.}}
\label{tab:top_perc}
\end{table}
% \begin{table}[t]
% \small
% \centering
% \begin{tabular}{l|r|r}
% \hline
% \textbf{Country} & \textbf{\# Violators} & \textbf{\% VPs}\\
% \toprule
% Eswatini & 1 & 100.00\% \\ %\ltgrey
% \rowcolor{gray!10}
% Lesotho & 1 & 100.00\% \\ 
% Mozambique & 2 & 66.67\% \\ %\ltgrey
% \rowcolor{gray!10}
% Botswana & 2 & 66.67\% \\
% Zambia & 1 & 14.29\% \\ %\ltgrey
% \rowcolor{gray!10}
% Azerbaijan & 1 & 14.29\% \\
% Armenia & 1 & 7.14\% \\ %\ltgrey
% \rowcolor{gray!10}
% Iceland & 1 & 7.14\% \\
% United Arab Emirates & 1 & 5.00\% \\ %\ltgrey
% \rowcolor{gray!10}
% Réunion & 2 & 4.88\%\\

% % ZA & 6 & 5.77\% \\
% % ID & 5 & 5.21\% \\
% \bottomrule
% \end{tabular}
% \caption{Geographic Skew of Misreported Probes---\textnormal{The majority of countries with the largest fraction of misreported probes in May~2024 are near South Africa: Eswatini, Lesotho, Mozambique, Botswana, Zambia, Réunion.}} %below the equator
% \label{tab:top_perc}
% \end{table}
%\vskip 0.1cm
% \vspace{-0.5cm}
\subsection{Distribution of Violating Probes}
Violating probes exist all over the world. 
In Table~\ref{tab:top_raw}, we list the countries with the most violating probes, nearly half of which originate in Germany (32\%) or the USA (17\%).
Within the US, violating probes are broadly distributed.
We infer that violating probes in Kansas are reported to be more than 200~km from the geographic center of the US~\cite{geocenter}, suggesting that the geolocation is likely misreported, not that RIPE automatically used the default US location.
% Within the US, the most (26\%) violating probes are located in Kansas, causing 10\% of probes hosted in Kansas to violate \sol.
% Additionally, within the US, 50\% (2/4) of Alaska's probes and 67\% (6/9) of Kansas' probes also violate the \sol.
% Thus, any study that relies on Alaska or Kansas probes is more likely than not using a violating probe. 

Countries in southern Africa are most likely to host at least one violating probe. 
Table~\ref{tab:top_perc} lists the top 10~countries with the highest proportion of probes violating the \sol. 
Many of the top~10 countries are in southern Africa: Botswana (100\% of probes violated), Lesotho (100\%), Mozambique (33.33\%), and Zambia (25\%).
Notably, the southern African countries have extremely low vantage point coverage to begin with (i.e., only one to four probes each), thereby exacerbating the effects of a violating probe.

Violating probes in southern Africa exhibit varying degrees of geolocation error. 
A probe in Botswana, for example, violates \sol by 2.5~ms, yet shows an RTT of just 0.33~ms to an \ark vantage point in Johannesburg—suggesting it lies within ~32~km of the city, excluding Botswana. 
In the worst case, a probe in Zambia violates \sol by 10.8~ms, but has a 1.16~ms RTT to Johannesburg, implying a location within ~160~km—still inconsistent with being in Zambia.
Both cases indicate that the infrastructure responding on behalf of the probes (e.g., the probe itself or a middlebox) are not in their reported countries. 
By not filtering for these regions in geolocation-dependent studies, we risk invalid conclusions. %, so future work should use our methodology to sanity check that the foundation of their study is properly evaluated.
% With plenty of prior work~\cite{surr_by_clouds,NREN_africa,fanou2017four} using RIPE Atlas probes to study African Internet connectivity, these \sol violations indicate the need to reproduce prior work's findings. 

The majority of violating probes respond from at least 1,600~km from their operator-reported geolocation. 
For every <probe, vantage point> pair that violates the SOI constraint, we calculate the minimum error of distance by 
(1) selecting all VPs that violated SOI,
(2) multiplying the minimum observed RTT by SOI, to derive the furthest distance, $d$, a probe could be if it traveled at exactly SOI,
(3) calculate the theoretical distance, $d_{theory}$, between our VP and the operator-reported probe location,
(4) subtract $d$ from $d_{theory}$, recording the minimum distance error. 
%a maximum bound on true distance between 
%multiplying recorded RTTs by the SOI to derive a maximum bound on the true distance between two vantage points. 
%We then subtract this theoretical distance from the metadata-reported distance to all 7 RIPE servers, recording the maximum difference of this subtraction.
We analyze the full distribution of distance errors: 80\% exceed 160~km, and 20\% are greater than 4,800~km (see Appendix Figure~\ref{fig:cdf_error}).
The large distance errors suggest violating probes are likely on different continents, as we discuss in Section~\ref{subsub:case}.
%We find no significant differences between violating probes and non-violating probes when comparing the network type they reside in and their age. 
%Autonomous System distributions and probe age. We find no significant difference between the makeup of the ASes announcing violating probes versus non-violating probes (e.g., the vast majority of ASes are ISPs and cloud providers~\cite{ziv2021asdb} for both). Violating probes are also not significantly older or younger than non-violating probes.

\subsection{Violating Probe Behavior Over Time}
\begin{figure}[t]
    \centering
    \includegraphics[width=\columnwidth]{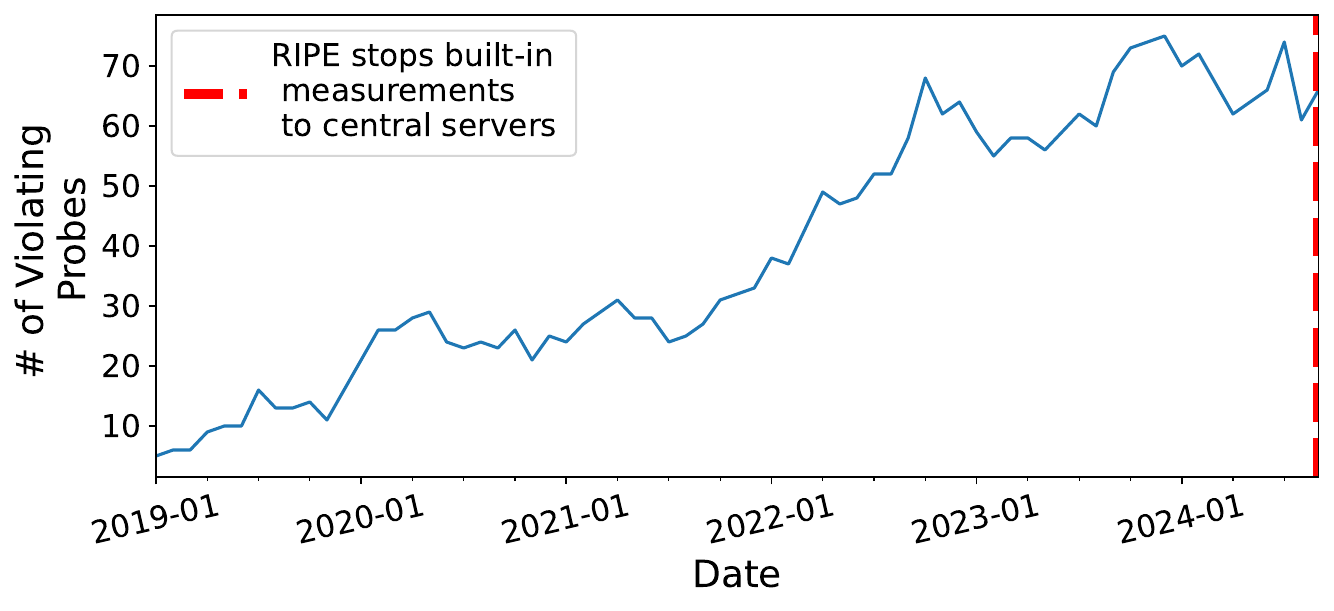}
    \caption{Violations over Time---\textnormal{Based on RIPE measurements alone, violating probes increased at least tenfold—from 5 probes in January 2019 to 66 probes in September 2024 (historical Atlas data is unavailable for later periods).
}}
    \label{fig:over_time}
\end{figure}
Across five~years, the raw number of probes that violate SOI increases every year.
Using historical data from built-in RIPE measurements, Figure~\ref{fig:over_time} illustrates a tenfold increase in likely violating probes, from just 5 in 2019 to 66 in September 2024.
%\footnote{We do not consider Ark measurements as they do not store historical data.}
Using both RIPE's historical measurement data from the past 5 years and our measurements with \ark from May~2024 to March~2025, we find a total of 664 violating probes (1.5\% of all probes from the past 5 years)
%\footnote{Notably, the fraction of violating probes found over 5~years is smaller than in May~2024 because we are unable to use our \ark measurement platform to verify latency back in time. Thus, the majority of these violating probes are found using only historical, RIPE Atlas built-in measurements.\label{no_ark_mzrments}} 
that respond from a different geolocation at some point in time.
Roughly 76\% (505/664) eventually stop violating SOI, with 10.5\% of those eventually becoming disconnected or abandoned after the SOI is violated.

%nearly 400 probes reported the wrong loc for at least 1 timeframe
%roughly 50% stop viol SOI, but 24% of these eventually just disconect or become abandoned after SOI violated
%^that has to be the last sentiment so we can transition to 76%

% Across 5~years, only half of violating probes stop violating.
% Using both RIPE's historical measurement data from the past 5 years and our present-day measurements from \ark, we find 391 violating probes (1\% of all probes from the past 5 years)\footnote{Notably, the fraction of violating probes found over 5~years is smaller than in May~2024 because we are unable to use our \ark measurement platform to verify latency back in time. Thus, the majority of these violating probes are found using only historical, RIPE Atlas built-in measurements.} that have reported the wrong location for at least one timeframe.
% Roughly 50\% (196/391) eventually stop violating SOI, with  24\% of those eventually becoming disconnected or abandoned after the SOI is violated.
%\todo{integrate over time fig~\ref{fig:over_time}}

The majority of probes that stop violating the SOI do so because the operator updates the probe's geolocation.
%\footnote{Although we cannot say whether the geolocation is updated to the correct location, the probe stops violating the SOI.}.
While 20\% of violating probes are updated in less than one week, the majority of violating probes take at least 7~weeks (Appendix Figure~\ref{fig:late}). 
We further find that 80\% of operators whose probes no longer violate SOI updated their metadata over 1,600~km, indicating that their probes could have incorrectly reported locations on different \textit{continents} (Appendix Figure~\ref{fig:loc_ch}).

\subsection{Understanding Contributing Factors}
%\todo{``Currently, the specific discussion of individual misreporting probes has no clear takeaway"}
Through a manual analysis of reported geolocations and round trip times, we discover that tardiness in updating geolocation after a move, and initial misreporting of geolocation are likely contributors to SOI violations.

\subsubsection{Tardiness} \label{subsub:case}
\begin{figure}[t]
    \centering
\includegraphics[width=\columnwidth]{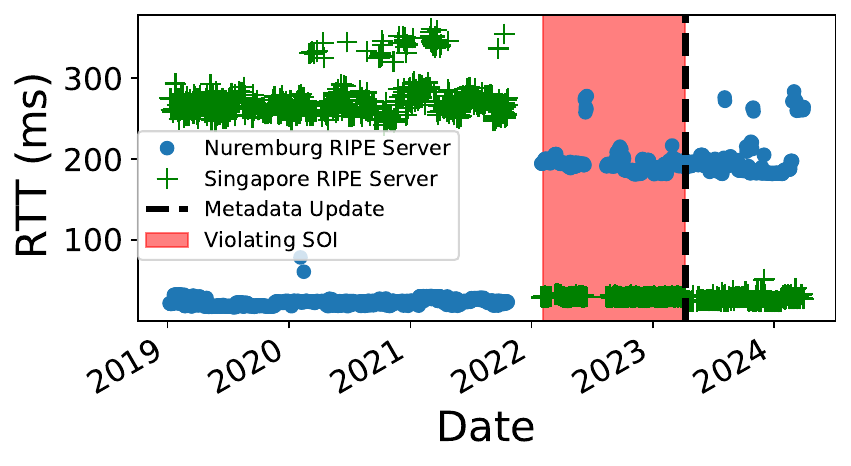}
    \caption{Moving from Germany to Brunei Case Study---\textnormal{The operator's probe reports a SOI violation between February 2022 and April 2023, before updating their location metadata from Germany to Brunei in April 2023.}}
    \label{fig:case_study}
\end{figure}
\vskip 0.2cm
At times, probes reporting RTTs that violate SOI are simply late to report a move.
Figure~\ref{fig:case_study} shows longitudinal ping measurements from probe 822 whose operator reported a location in Germany. 
From 2019 to 2022, its RTTs to the Nuremberg RIPE servers were a minimum of 17.2 ms, indicating the probe was likely within 1,700~km of Nuremberg. 
After a disconnect between November~2021 and February~2022, the probe reported a minimum RTT of 23~ms to the Singapore RIPE server, indicating it was likely within 2,400~km of Singapore. However, until April 2023, the probe reported a location in Germany, which is not within 2,400~km of Singapore. 
In April 2023, the operator updated their metadata to Brunei, which is within 2,400~km of Singapore. 
The probe no longer reports RTTs in violation of the SOI.

%While a location change may be the most intuitive explanation of momentary SOI violations, we also show that a violation of SOI may instead be a result of operators incorrectly reporting metadata from the beginning of deployment.

\subsubsection{Initial Misreporting.}
%Operators accidentally misreporting a probe's geolocation contributes to long-term violations. 
%In other scenarios, probes misreporting geolocation from the beginning contributes to long-term violations.

Some long-term violations stem from probes configured with incorrect geolocation metadata. 
We examine Probe 1000011, which violated \sol from October 2019 to June 2023.
It reported a location in Marina Del Ray, CA until June 2023, when it switched to Arlington, VA. However, its minimum RTT to all RIPE servers remained stable---unexpected for a cross-country move. 
Moreover, RTTs to both Fremont and Newark RIPE servers violated \sol under the Marina Del Ray location, but no longer did once the location was updated. This suggests the probe was likely near New Jersey all along from 2019 until 2023. 
The location change coincides with an upstream provider switch from Cogent to Lumen, hinting that a network change may have prompted the operator to verify the probe's configuration and update the geolocation.

\subsection{Evaluating Detection Methodologies}
\label{sub:sec:eval}
We evaluate our methodology (Section~\ref{sec:meth}), by analyzing what biases are introduced when using different vantage points and how its performance compares to prior work. 
Our methodology uses an order of magnitude less bandwidth and time. 
%We show how a moderate amount of ground truth---namely, a set of 157 verified vantage points---can find 3~times more violating probes than using only RIPE infrastructure, while using nearly an order of magnitude less bandwidth and less time. 
\begin{figure}
    \centering
    \includegraphics[width=\columnwidth]{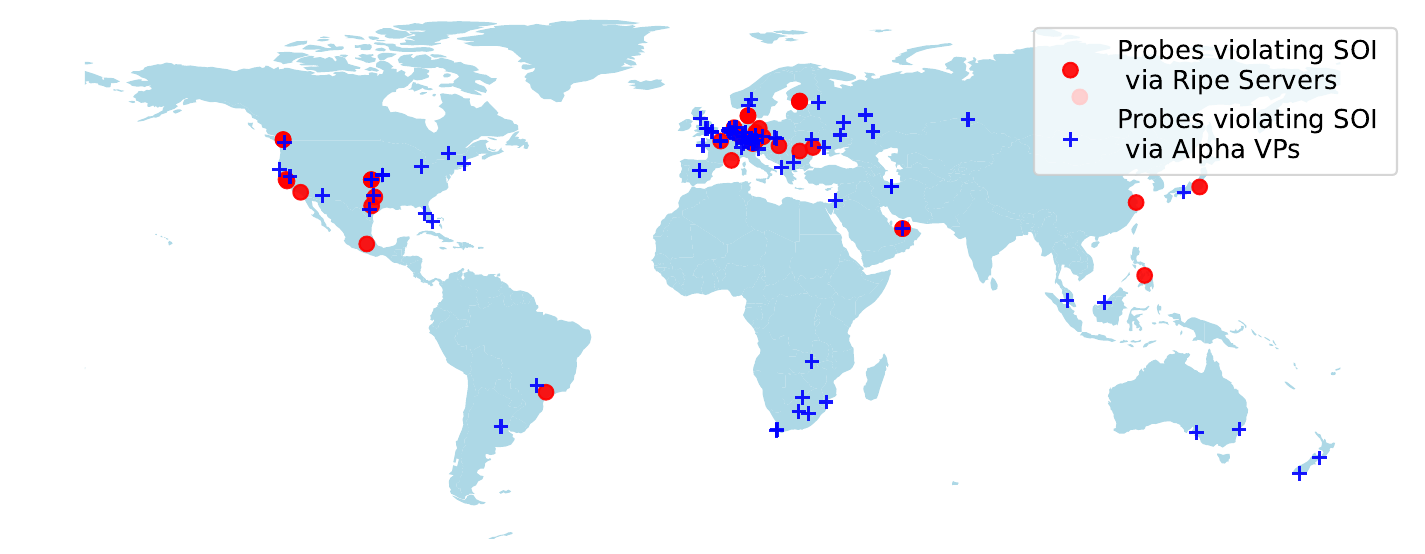}
        \caption{Coverage of Violations---\textnormal{RIPE servers suffice to verify SOI violations in Western and Central Europe, but studies in southern Africa, Southeast Asia, Australia, and Eastern Europe should also use \ark.}}
    \label{fig:ark_ripe_cov}
\end{figure}
\subsubsection{Impact of Vantage Points}
While coupling RIPE and \ark to detect SOI violations maximizes finding violating probes, certain geographical regions benefit from using \ark more than others. In Figure~\ref{fig:ark_ripe_cov}, we plot all probes found violating SOI today via RIPE servers and all probes found violating SOI today via \ark vantage points. Due to the location of \ark servers, \ark provides more opportunities to find SOI violations in southern Africa, Southeast Asia, Australia, and Eastern Europe.

\subsubsection{Comparison to Prior Work}

\todo{remove and massage related work accordingly}
We show that existing methodologies require significantly more bandwidth and flag significantly more false positives than our methodology.  
%In comparison, our methodology remains effective and efficient. 

\label{sub:sub:sec:comp_prior}
% Prior work uses RIPE Anchors instead of RIPE probes to validate SOI violations, which increases the density of vantage points considerably (157 versus 1300), but risks using invalidated infrastructure. As introduced in Section~\ref{sub:sec:research_related}, Darwich et al. develop a methodology to filter violating probes using Anchors.

\vspace{3pt}
\noindent
\textbf{Darwich et al.~\cite{darwich2023replication} Comparison.}\quad
%We run Darwich et al.~\cite{darwich2023replication}'s open-source methodology to identify probes that violate the SOI on data collected at the time of their study. 
Darwich et al. propose a three-step methodology: (1) all \atlas anchors ping each other, (2) iteratively remove anchors with the most \sol violations until none remain, and (3) test all probes against the validated anchors to identify violations.

% Darwich et al. propose a methodology that consists of
% (1) \atlas anchors all pinging each other, 
% (2) iteratively removing anchors with the largest number of SOI violations until no more anchors violate the SOI, and
% (3) running ping tests from every probe to the set of validated anchors (step 2) to produce a list of probes that violate the SOI. 

Our methodology runs in an order of magnitude less time and uses nearly an order of magnitude less bandwidth.
Darwich et al. state their methodology runs between several hours to several days (bottlenecked by \atlas measurement limits) to collect 9M ping measurements on the \atlas platform.
In comparison, our methodology---pulling historical data from RIPE's API and running \ark measurements---takes only 18 minutes to collect 1.6M ping measurements.
Darwich et al.'s methodology uses ``millions of RIPE credits,'' deploying $1300\times 
1300$ iterative measurements for the anchors, plus the $1300\times 13K$ iterative measurements for the probes at one point in time. 
Our methodology uses substantially fewer (verified) vantage points, requiring 9~times less ($157 \times 13K$) iterative measurements for one point in time.

Unfortunately, we cannot compare the number of violating probes Darwich et al.'s methodology finds compared to ours. 
At the time of their study, Darwich et al.'s methodology finds 105~violating probes. 
However, our methodology cannot be applied to the time of their study, as we do not have historical data for the \ark platform.
In May~2024, our methodology finds 197~violating probes. 
However, replicating their study requires millions of RIPE credits that we simply do not have access to (and by default are not given to researchers). 
While their methodology does not find violating probes in low-coverage areas (e.g., Lesotho), it is not clear if this is due to a temporal change. 
%\vskip 0.5cm
%Darwich et al.'s methodology is infeasible to reproduce as it requires hundreds of millions of credits. To acquire 100 million credits, one would either need to host a probe for 12.68 years, host an anchor 1.27 years, or become a platinum sponsor (costing upwards of €100,000) ~\cite{ripeAtlasCredits, ripeAtlasSponsor}.

\vspace{3pt}
\noindent
\textbf{Gharaibeh et al.~\cite{gharaibeh2017look} Comparison.}\quad
Gharaibeh et al.
consider geolocations untrustworthy if (1) they report default country coordinates (typically geographic centers) or 
(2) two probes are connected to the same router, but report locations more than 100km apart.
We replicate their methodology on built-in \atlas traceroutes and find that it flags roughly one quarter of probes as likely having misreported geolocation.
Upon investigation, we discover that approximately 3K probes share the same router in their traceroute, causing the entire group to be considered untrustworthy.
Within that group is \atlas' own probe (probe \#1), which \atlas confirms to us is reporting the correct geolocation. 
% Gharaibeh et al.
% consider reported geolocations untrustworthy if (1) they report default country coordinates (typically geographic centers) or 
% (2) two probes are connected to the same router, but report locations more than 100km apart.
% We replicate their methodology on built-in \atlas traceroutes and find that it flags roughly one quarter of probes as likely having misreported geolocation.
% Upon investigation, we discover that \changed{80 probes were reported to be less than 5km from their geographic center~\cite{geocenter}. We manually conducted traceroute measurements to these probes and found that most of the responsive probes had reasonable locations (e.g. the reverse DNS record of a hop IP close to the destination included a nearby city name abbreviation). Two probes that were incorrectly reported as being in the center of the U.K. were also flagged by our methodology. For other probes located in small countries, such as Singapore, it is plausible that they were indeed located near the geographic center of those countries. We also analyzed the routers to which the probes were connected and found} roughly 3k probes share the same router in their traceroute, causing the entire group to be considered untrustworthy.
% Within that group is \atlas' own probe (probe \#1), which \atlas confirms to us is indeed reporting the correct geolocation. 
\vskip 0.5cm
    \section{Limitations}\label{sec:limitations}
Our work faces several limitations in providing an in-depth understanding of operator-misreported geolocation. 
First, we focus on the \atlas dataset, which may not be representative of all vantage points with operator-reported geolocations.
It is critical that future work verify how the prevalence of violating probes affects other vantage point platforms. 
Nevertheless, \atlas is the largest community-vantage point collection, underscoring the importance of our results. 
Second, our method cannot distinguish between the probe's true location and that of a device responding on its behalf (e.g., a middlebox). However, since many geolocation methods rely on latency to the responder, it’s the responder’s location that ultimately matters. Third, our method produces a strict lower bound by requiring vantage points to be slower than ideal, congestion-free links—so the true number of violations is likely even higher.
\vskip 0.5cm
% Thus, our work could be severely underestimating the prevalence of violating probes and their distance error. 

    \section{Conclusion}

As operator-reported geolocations continue to be used as a foundation for many research studies, it is critical that we understand and maintain an accurate ground truth.
In this work, we provide an in-depth analysis of operator-\textit{mis}reported geolocations. 
Between May 2024 and March 2025, we conservatively infer that at least 470 (3.96\%) probes likely do not respond from their operator-reported geographic location.
However, misreported geolocations are often hundreds of kilometers away from their reported locations, leaving entire countries without the majority---if not all---of their probe coverage. 
Furthermore, the problem is only getting worse: within the past five years, the number of probes reporting the wrong location has increased tenfold. 
We find that operator error is often at fault, with 76\% of probes that historically violated SOI no longer violating SOI due to updated metadata.
Furthermore, it takes most operators over 7 weeks to update metadata after SOI violations begin. 

We open-source our methodology, release a list of 664 likely inaccurately geolocated RIPE Atlas probes from the past five years, and maintain a weekly-updated list of currently violating \atlas probes
to help researchers sanity check the dataset they rely upon. 
%Our filtering methodology is efficient and effective, so we recommend that anyone using \atlas's operator-reported geolocation should run our methodology and verify that the geolocation of a probe's response is not physically impossible.
%To increase coverage, researchers should use our \ark platform to perform our SOI methodology to minimize overall time and bandwidth. 
Nevertheless, future work should explore new methods to detect SOI violations, including providing tighter---but still accurate---bounds for detecting violations. 
We also recommend that \atlas extend their current communication infrastructure to account for SOI violations. RIPE Atlas currently emails operators when their probe disconnects; therefore emailing operators when a reported location \textit{seems} false by physical standards could be a natural extension to an existing system. By doing so, RIPE Atlas could minimize the bandwidth researchers would use to detect the violations on their own and could perhaps deploy more stationary probes in the regions now lacking coverage. 
\looseness=-1

% For now, we are currently in contact with \atlas to resolve the SOI violations affecting 196~probes as of April~2024. 
%\vskip 0.5cm
% itself run \sol violation tests.
% \atlas can email operators to implore them to update their geolocation, to reduce the median 7~week overhead it takes an operator to change geolocation.
% \atlas should also flag, and/or disconnect probes that violate \sol until the operator resolves the issue, to minimize the bandwidth researchers will use to detect the violations on their own. 

% More generally, our community should be more vigilant when using crowd-sourced data, especially in sensitive settings such as measuring latency for geolocation. 
% Internet measurement researchers in the past have also found that crowd-sourcing leads to high variance in accuracy in other settings, sometimes due to lack of incentive and other times due to natural human error~\cite{ziv2021asdb}.

%RIPE Atlas is the largest crowd-sourced measurement platform, so should another measurement platform rise, this methodology can also be applied. Other future work could develop new methods to detect SOI violations when a probe has changed location, perhaps even predicting where the new location could be. 

%\keeping{recs for RIPE for community, how much the problem persists and how long it takes people to upodate their locstion. there is a diff btw where a server is and where the thing that responds actually is. as a community, when we say we want to geolocate things, do we mean the actual thing or what responds instead?}

    \bibliographystyle{ACM-Reference-Format}
    \bibliography{refs}
    \clearpage

\fi

\ifietf
  
  \maketitle
  
  \input{ietf_workshop/meth}
  \input{ietf_workshop/res}
  \input{ietf_workshop/future}
  \bibliographystyle{ACM-Reference-Format}
\bibliography{refs}
    
\fi

% \bibliographystyle{ACM-Reference-Format}

% \bibliography{refs}
% \clearpage
% \input{appendix.tex}

\end{document}